\documentstyle[10pt,epsfig,dp_delphititle,float,hangcaption,xspace,amssymb,
amsfonts,amsmath,amsthm,cite,graphicx,lineno]{dp_delphi}
\makeindex
\def\DpPaperGroup{PH-EP}
\def\DpPaperRef{2004-017}
\def\DpDate{27 April 2004}
\def\DpAuthors{DELPHI Collaboration}
\def\DpSubmit{(Accepted by Eur. Phys. J. C)}
\def\DpTitle{{Determination of the $e^+e^- \to \gamma \gamma (\gamma)$ 
cross-section  at LEP 2 }}

\newcommand{\eeto} {e^+e^- \to}

\newcommand{\gr} {\mbox{$^{\circ}$}}

\newcommand{\fot} {\mbox{$\gamma$}}
\newcommand{\bfot} {\mbox{$\boldsymbol\gamma$}}

\newcommand{\scos} {\mbox{$\rm{\cos^2}$}}

\newcommand{\eeintogg} {\mbox{$\eeto \, \gamma \gamma$}}
\newcommand{\beeintogg}{\mbox{${\boldmath \boldsymbol e^+ \boldsymbol e^-
\boldsymbol \to \boldsymbol \gamma \boldsymbol \gamma}$}}

\newcommand{\auth}[1]{#1,}

\newcommand{\etal}{et al.}
\newcommand{\titl}[1]{{\it ``#1''}}

\newcommand{\PTF}[3]{{\sl Prog. of Theor. Phys.} {\bf Vol. #1} (#2)~#3}
\newcommand{\npB}[3]{{\sl Nucl. Phys.} {\bf B#1} (#2)~#3}
\newcommand{\PLB}[3]{{\sl Phys. Lett.} {\bf B#1} (#2)~#3}

\newcommand{\prD}[3]{{\sl Phys. Rev.} {\bf D#1} (#2) #3}
\newcommand{\prl}[3]{{\sl Phys. Rev. Lett.} {\bf #1} (#2) #3}

\newcommand{\zp}[3]{{\sl Zeit.\ Phys.} {\bf C#1} (#2) #3}

\newcommand{\cpc}[3]{{\sl Comp.\ Phys.\ Comm.} {\bf #1} (#2) #3}
\newcommand{\EPJ}[3]{{\sl Eur. Phys. J.} {\bf C#1} (#2) #3}
\newcommand{\nim}[3]{{\sl Nucl.\,Instr.\,and\,Meth.} {\bf A#1} (#2) #3}
\begin{document}
\makeatletter
\newcount\@tempcntc
\def\@citex[#1]#2{\if@filesw\immediate\write\@auxout{\string\citation{#2}}\fi
  \@tempcnta\z@\@tempcntb\m@ne\def\@citea{}\@cite{\@for\@citeb:=#2\do
    {\@ifundefined
       {b@\@citeb}{\@citeo\@tempcntb\m@ne\@citea\def\@citea{,}{\bf ?}\@warning
       {Citation `\@citeb' on page \thepage \space undefined}}%
    {\setbox\z@\hbox{\global\@tempcntc0\csname b@\@citeb\endcsname\relax}%
     \ifnum\@tempcntc=\z@ \@citeo\@tempcntb\m@ne
       \@citea\def\@citea{,}\hbox{\csname b@\@citeb\endcsname}%
     \else
      \advance\@tempcntb\@ne
      \ifnum\@tempcntb=\@tempcntc
      \else\advance\@tempcntb\m@ne\@citeo
      \@tempcnta\@tempcntc\@tempcntb\@tempcntc\fi\fi}}\@citeo}{#1}}
\def\@citeo{\ifnum\@tempcnta>\@tempcntb\else\@citea\def\@citea{,}%
  \ifnum\@tempcnta=\@tempcntb\the\@tempcnta\else
   {\advance\@tempcnta\@ne\ifnum\@tempcnta=\@tempcntb \else \def\@citea{--}\fi
    \advance\@tempcnta\m@ne\the\@tempcnta\@citea\the\@tempcntb}\fi\fi}
 
\makeatother
\begin{titlepage}
\pagenumbering{roman}
\CERNpreprint{\DpPaperGroup}{\DpPaperRef} 
\date{{\small\DpDate}} 
\title{\DpTitle} 
\address{\DpAuthors} 
\begin{shortabs} 
\noindent


A test of the benchmark QED process \eeintogg(\fot) is reported, 
using the data collected with the DELPHI detector at LEP 2. 
The data analysed were recorded at centre-of-mass energies 
ranging from 161~GeV to 208~GeV and correspond to a total 
integrated luminosity of 656.4~pb$^{-1}$.
The Born cross-section for the process \eeintogg(\fot) was 
determined, confirming the validity of QED at the highest 
energies ever attained in electron-positron collisions.
Lower limits on the parameters of 
a number of possible deviations from QED, predicted within theoretical frameworks expressing physics beyond the Standard Model, were derived.

\end{shortabs}
\vfill
\begin{center}
\DpSubmit \ \\ 
\end{center}
\vfill
\clearpage
\headsep 10.0pt
\addtolength{\textheight}{10mm}
\addtolength{\footskip}{-5mm}
\begingroup
%
\newcommand{\DpName}[2]{\hbox{#1$^{\ref{#2}}$},\hfill}
\newcommand{\DpNameTwo}[3]{\hbox{#1$^{\ref{#2},\ref{#3}}$},\hfill}
\newcommand{\DpNameThree}[4]{\hbox{#1$^{\ref{#2},\ref{#3},\ref{#4}}$},\hfill}
\newskip\Bigfill \Bigfill = 0pt plus 1000fill
\newcommand{\DpNameLast}[2]{\hbox{#1$^{\ref{#2}}$}\hspace{\Bigfill}}
%
\footnotesize
\noindent
\DpName{J.Abdallah}{LPNHE}
\DpName{P.Abreu}{LIP}
\DpName{W.Adam}{VIENNA}
\DpName{P.Adzic}{DEMOKRITOS}
\DpName{T.Albrecht}{KARLSRUHE}
\DpName{T.Alderweireld}{AIM}
\DpName{R.Alemany-Fernandez}{CERN}
\DpName{T.Allmendinger}{KARLSRUHE}
\DpName{P.P.Allport}{LIVERPOOL}
\DpName{U.Amaldi}{MILANO2}
% ...
\normalsize
\endgroup
\titlefoot{Department of Physics and Astronomy, Iowa State
     University, Ames IA 50011-3160, USA
    \label{AMES}}
\titlefoot{Physics Department, Universiteit Antwerpen,
     Universiteitsplein 1, B-2610 Antwerpen, Belgium \\
     \indent~~and IIHE, ULB-VUB,
     Pleinlaan 2, B-1050 Brussels, Belgium \\
     \indent~~and Facult\'e des Sciences,
     Univ. de l'Etat Mons, Av. Maistriau 19, B-7000 Mons, Belgium
    \label{AIM}}
\titlefoot{Physics Laboratory, University of Athens, Solonos Str.
     104, GR-10680 Athens, Greece
    \label{ATHENS}}
\titlefoot{Department of Physics, University of Bergen,
     All\'egaten 55, NO-5007 Bergen, Norway
    \label{BERGEN}}
\titlefoot{Dipartimento di Fisica, Universit\`a di Bologna and INFN,
     Via Irnerio 46, IT-40126 Bologna, Italy
    \label{BOLOGNA}}
\titlefoot{Centro Brasileiro de Pesquisas F\'{\i}sicas, rua Xavier Sigaud 150,
     BR-22290 Rio de Janeiro, Brazil \\
     \indent~~and Depto. de F\'{\i}sica, Pont. Univ. Cat\'olica,
     C.P. 38071 BR-22453 Rio de Janeiro, Brazil \\
     \indent~~and Inst. de F\'{\i}sica, Univ. Estadual do Rio de Janeiro,
     rua S\~{a}o Francisco Xavier 524, Rio de Janeiro, Brazil
    \label{BRASIL}}
\titlefoot{Coll\`ege de France, Lab. de Physique Corpusculaire, IN2P3-CNRS,
     FR-75231 Paris Cedex 05, France
    \label{CDF}}
\titlefoot{CERN, CH-1211 Geneva 23, Switzerland
    \label{CERN}}
\titlefoot{Institut de Recherches Subatomiques, IN2P3 - CNRS/ULP - BP20,
     FR-67037 Strasbourg Cedex, France
    \label{CRN}}
\titlefoot{Now at DESY-Zeuthen, Platanenallee 6, D-15735 Zeuthen, Germany
    \label{DESY}}
\titlefoot{Institute of Nuclear Physics, N.C.S.R. Demokritos,
     P.O. Box 60228, GR-15310 Athens, Greece
    \label{DEMOKRITOS}}
\titlefoot{FZU, Inst. of Phys. of the C.A.S. High Energy Physics Division,
     Na Slovance 2, CZ-180 40, Praha 8, Czech Republic
    \label{FZU}}
\titlefoot{Dipartimento di Fisica, Universit\`a di Genova and INFN,
     Via Dodecaneso 33, IT-16146 Genova, Italy
    \label{GENOVA}}
\titlefoot{Institut des Sciences Nucl\'eaires, IN2P3-CNRS, Universit\'e
     de Grenoble 1, FR-38026 Grenoble Cedex, France
    \label{GRENOBLE}}
\titlefoot{Helsinki Institute of Physics, P.O. Box 64,
     FIN-00014 University of Helsinki, Finland
    \label{HELSINKI}}
\titlefoot{Joint Institute for Nuclear Research, Dubna, Head Post
     Office, P.O. Box 79, RU-101 000 Moscow, Russian Federation
    \label{JINR}}
\titlefoot{Institut f\"ur Experimentelle Kernphysik,
     Universit\"at Karlsruhe, Postfach 6980, DE-76128 Karlsruhe,
     Germany
    \label{KARLSRUHE}}
\titlefoot{Institute of Nuclear Physics PAN,Ul. Radzikowskiego 152,
     PL-31142 Krakow, Poland
    \label{KRAKOW1}}
\titlefoot{Faculty of Physics and Nuclear Techniques, University of Mining
     and Metallurgy, PL-30055 Krakow, Poland
    \label{KRAKOW2}}
\titlefoot{Universit\'e de Paris-Sud, Lab. de l'Acc\'el\'erateur
     Lin\'eaire, IN2P3-CNRS, B\^{a}t. 200, FR-91405 Orsay Cedex, France
    \label{LAL}}
\titlefoot{School of Physics and Chemistry, University of Lancaster,
     Lancaster LA1 4YB, UK
    \label{LANCASTER}}
\titlefoot{LIP, IST, FCUL - Av. Elias Garcia, 14-$1^{o}$,
     PT-1000 Lisboa Codex, Portugal
    \label{LIP}}
\titlefoot{Department of Physics, University of Liverpool, P.O.
     Box 147, Liverpool L69 3BX, UK
    \label{LIVERPOOL}}
\titlefoot{Dept. of Physics and Astronomy, Kelvin Building,
     University of Glasgow, Glasgow G12 8QQ
    \label{GLASGOW}}
\titlefoot{LPNHE, IN2P3-CNRS, Univ.~Paris VI et VII, Tour 33 (RdC),
     4 place Jussieu, FR-75252 Paris Cedex 05, France
    \label{LPNHE}}
\titlefoot{Department of Physics, University of Lund,
     S\"olvegatan 14, SE-223 63 Lund, Sweden
    \label{LUND}}
\titlefoot{Universit\'e Claude Bernard de Lyon, IPNL, IN2P3-CNRS,
     FR-69622 Villeurbanne Cedex, France
    \label{LYON}}
\titlefoot{Dipartimento di Fisica, Universit\`a di Milano and INFN-MILANO,
     Via Celoria 16, IT-20133 Milan, Italy
    \label{MILANO}}
\titlefoot{Dipartimento di Fisica, Univ. di Milano-Bicocca and
     INFN-MILANO, Piazza della Scienza 2, IT-20126 Milan, Italy
    \label{MILANO2}}
\titlefoot{IPNP of MFF, Charles Univ., Areal MFF,
     V Holesovickach 2, CZ-180 00, Praha 8, Czech Republic
    \label{NC}}
\titlefoot{NIKHEF, Postbus 41882, NL-1009 DB
     Amsterdam, The Netherlands
    \label{NIKHEF}}
\titlefoot{National Technical University, Physics Department,
     Zografou Campus, GR-15773 Athens, Greece
    \label{NTU-ATHENS}}
\titlefoot{Physics Department, University of Oslo, Blindern,
     NO-0316 Oslo, Norway
    \label{OSLO}}
\titlefoot{Dpto. Fisica, Univ. Oviedo, Avda. Calvo Sotelo
     s/n, ES-33007 Oviedo, Spain
    \label{OVIEDO}}
\titlefoot{Department of Physics, University of Oxford,
     Keble Road, Oxford OX1 3RH, UK
    \label{OXFORD}}
\titlefoot{Dipartimento di Fisica, Universit\`a di Padova and
     INFN, Via Marzolo 8, IT-35131 Padua, Italy
    \label{PADOVA}}
\titlefoot{Rutherford Appleton Laboratory, Chilton, Didcot
     OX11 OQX, UK
    \label{RAL}}
\titlefoot{Dipartimento di Fisica, Universit\`a di Roma II and
     INFN, Tor Vergata, IT-00173 Rome, Italy
    \label{ROMA2}}
\titlefoot{Dipartimento di Fisica, Universit\`a di Roma III and
     INFN, Via della Vasca Navale 84, IT-00146 Rome, Italy
    \label{ROMA3}}
\titlefoot{DAPNIA/Service de Physique des Particules,
     CEA-Saclay, FR-91191 Gif-sur-Yvette Cedex, France
    \label{SACLAY}}
\titlefoot{Instituto de Fisica de Cantabria (CSIC-UC), Avda.
     los Castros s/n, ES-39006 Santander, Spain
    \label{SANTANDER}}
\titlefoot{Inst. for High Energy Physics, Serpukov
     P.O. Box 35, Protvino, (Moscow Region), Russian Federation
    \label{SERPUKHOV}}
\titlefoot{J. Stefan Institute, Jamova 39, SI-1000 Ljubljana, Slovenia
     and Laboratory for Astroparticle Physics,\\
     \indent~~Nova Gorica Polytechnic, Kostanjeviska 16a, SI-5000 Nova Gorica, Slovenia, \\
     \indent~~and Department of Physics, University of Ljubljana,
     SI-1000 Ljubljana, Slovenia
    \label{SLOVENIJA}}
\titlefoot{Fysikum, Stockholm University,
     Box 6730, SE-113 85 Stockholm, Sweden
    \label{STOCKHOLM}}
\titlefoot{Dipartimento di Fisica Sperimentale, Universit\`a di
     Torino and INFN, Via P. Giuria 1, IT-10125 Turin, Italy
    \label{TORINO}}
\titlefoot{INFN,Sezione di Torino, and Dipartimento di Fisica Teorica,
     Universit\`a di Torino, Via P. Giuria 1,\\
     \indent~~IT-10125 Turin, Italy
    \label{TORINOTH}}
\titlefoot{Dipartimento di Fisica, Universit\`a di Trieste and
     INFN, Via A. Valerio 2, IT-34127 Trieste, Italy \\
     \indent~~and Istituto di Fisica, Universit\`a di Udine,
     IT-33100 Udine, Italy
    \label{TU}}
\titlefoot{Univ. Federal do Rio de Janeiro, C.P. 68528
     Cidade Univ., Ilha do Fund\~ao
     BR-21945-970 Rio de Janeiro, Brazil
    \label{UFRJ}}
\titlefoot{Department of Radiation Sciences, University of
     Uppsala, P.O. Box 535, SE-751 21 Uppsala, Sweden
    \label{UPPSALA}}
\titlefoot{IFIC, Valencia-CSIC, and D.F.A.M.N., U. de Valencia,
     Avda. Dr. Moliner 50, ES-46100 Burjassot (Valencia), Spain
    \label{VALENCIA}}
\titlefoot{Institut f\"ur Hochenergiephysik, \"Osterr. Akad.
     d. Wissensch., Nikolsdorfergasse 18, AT-1050 Vienna, Austria
    \label{VIENNA}}
\titlefoot{Inst. Nuclear Studies and University of Warsaw, Ul.
     Hoza 69, PL-00681 Warsaw, Poland
    \label{WARSZAWA}}
\titlefoot{Fachbereich Physik, University of Wuppertal, Postfach
     100 127, DE-42097 Wuppertal, Germany
    \label{WUPPERTAL}}
\addtolength{\textheight}{-10mm}
\addtolength{\footskip}{5mm}
\clearpage
\headsep 30.0pt
\end{titlepage}
%
\pagenumbering{arabic} 
\setcounter{footnote}{0} %
\large
\section{Introduction}

An analysis of the process \eeintogg(\fot) is reported.
The data analysed were collected with the DELPHI detector \cite{bib:delphi} at LEP~2, 
at collision energies, $\sqrt{s}$, ranging from 161~GeV up to~208~GeV, 
corresponding to a total integrated luminosity of 656.4~pb$^{-1}$.
The studied reaction is an almost pure QED (Quantum ElectroDynamics) process 
which, at orders above $\alpha^2$, is mainly affected 
by QED corrections, such as soft and hard {\it bremsstrahlung}
and virtual corrections, compared to which the  
weak corrections due to the exchange of virtual massive gauge bosons are very small \cite{bib:BerKl,bib:bohm,bib:fuji}.
Therefore, any significant deviation between the measured and the predicted QED 
cross-sections could unambiguously be interpreted as a manifestation of 
non-standard physics.
Moreover, the differential cross-section terms 
expressing interferences between QED and various non-standard physics processes, 
behave very differently from the QED term in their 
dependence on the scattering angle of the photons with respect to the 
incident electron/positron. 
Depending on the possible new physics scenario, a 
departure of the measured differential cross-section of \eeintogg\ 
from the Standard Model expectations, could then be interpreted as a
measure of the energy scale of the QED breakdown~\cite{bib:drell,bib:low},  
of the characteristic energy scales of $e^+e^-\gamma\gamma$ contact interactions~\cite{bib:eboli}, of the mass of excited electrons within composite models~\cite{bib:vachon} 
or of the string mass scale~\cite{bib:giudice,bib:agashe} (which could be of the order of 
the electroweak scale in the framework of models with gravity propagating in 
large extra-dimensions).

The data recorded by DELPHI at LEP~2 were treated 
according to common reconstruction procedures, selection criteria and treatment of 
systematic uncertainties.
Previous results concerning the process $e^+e^- \to \gamma \gamma (\gamma)$, 
using DELPHI LEP~1 data are reported in reference \cite{bib:94}. 
An analysis of the LEP~2 data collected in 1996 and 1997 can be found in reference \cite{bib:96-97}, 
while the analysis of the 1998 and 1999 DELPHI data, from which the present analysis framework evolved is reported in \cite{bib:98-99}.
The results reported in the previous publications concerning the analyses of LEP 2 data 
are superseded by the results hereby reported. 
The most recently published results from the other LEP collaborations
can be found in references \cite{bib:aleph,bib:l3,bib:opal}.

\section{Data sample and apparatus}

The data analysed were recorded with the DELPHI detector at LEP 2 
from 1996 through 2000. 
They were grouped in ten subsets, as listed in table \ref{tab:data}, 
according to their centre-of-mass energy value or, 
in the case of the year 2000 data, split in two sets corresponding to different 
data processings (reconstruction procedures). This was made necessary due to 
an irreversible failure in a sector of the Time Projection Chamber, one of the most 
important tracking detectors, during the last part of the data taking period.

Events were generated for all the processes at the different 
centre-of-mass energies and passed through 
the full DELPHI simulation and reconstruction chain.
The event generator used to simulate the QED process $e^+e^-\to \gamma\gamma(\gamma)$ 
was that of reference~\cite{bib:BerKl} while Bhabha ($e^+e^- \to e^+e^- (\gamma)$) and 
Compton ($e^\pm \gamma$) 
scattering\footnote{Compton events are produced when a beam electron is scattered off  
by a quasi-real photon radiated by the other incoming electron, resulting mostly in final states with one photon and one visible electron, the remaining $e^\pm$ being lost in the beam pipe.} 
backgrounds were simulated with the BHWIDE~\cite{bib:bhwide} and TEEG~\cite{bib:compton} 
generators and $e^+e^- \to f \bar{f} (\gamma) $ events, including $e^+e^- \to \nu \bar{\nu} \gamma \gamma$ events, were generated with KORALZ~\cite{bib:KZ}. Two photon collision events,
yielding $e^+e^- f\bar{f}$ final states, were generated with BDK/BDKRC ~\cite{bib:bdk}.

The luminosity corresponding to the data sets analysed was measured 
by counting the number of Bhabha events at small polar angles~\cite{bib:STIC}.
These were recorded with the Small angle TIle Calorimeter (STIC)
that consisted of two cylindrical electromagnetic calorimeters surrounding the beam pipe 
at a distance of $\pm$220~cm from the interaction point, covering the 
polar angle\footnote{The polar angle, $\theta$, was defined with respect 
to the direction of the electron beam.} 
range from 2\gr\ to 10\gr\ and from 170\gr\ to 178\gr.

Photon detection and reconstruction relied on the trigger 
and energy measurements provided by the electro\-magnetic 
calorimeters: the STIC, the High density Projection Chamber (HPC) 
in the barrel region and the Forward ElectroMagnetic Calorimeter (FEMC) in the endcaps.
The HPC was a gas-sampling calorimeter, composed of 144 modules, each with 
9 longitudinal samplings in 41 lead layers and a scintillator layer 
which provided first and second level trigger signals.
It covered polar angles between 42\gr\ and 138\gr. 
The FEMC was a lead glass calorimeter, 
covering the polar angle regions $[11^\circ,35^\circ]$
and $[145^\circ,169^\circ]$.
The barrel DELPHI electro\-magne\-tic trigger~\cite{bib:trigger} 
required coincidences between scintillator signals 
and energy deposits in the HPC while in the forward 
region the electromagnetic trigger was given 
by energy deposits in the FEMC lead-glass counters.

The tracking system allowed the rejection of charged particles and the
recovery of photons converting inside the detector.
The DELPHI barrel tracking system relied on the Vertex Detector (VD), 
the Inner Detector (ID), the Time Projection Chamber (TPC) and the 
Outer Detector (OD).
In the endcaps, the tracking system relied also on the VD and the TPC
(down to about 20\gr\ in polar angle),
and on the Forward Chambers A and B (FCA, FCB).
The VD played an important role in the
detection of charged particle tracks coming from the interaction
point.

A more detailed description of the DELPHI detector, of the 
triggering conditions and of the readout chain can be found 
in \cite{bib:delphi,bib:trigger}.

\section{Event selection}
 

Preselection requirements enabled the rejection of most 
final states not compatible with multi-photon events.
Energy deposits in the HPC, FEMC and HCAL were used if their energy was greater than
500~MeV, 400~MeV and 900~MeV respectively.
The multiplicity of well reconstructed charged particle tracks 
was required to be less than 6, where a well reconstructed charged
particle track had to have momentum above 200~MeV/c, $z$ and $r\phi$ 
impact parameters below 4~cm/$\sin\theta$ and 4~cm respectively,
$\Delta p/p$ below 1.5 and could not have left signals in the VD only.
Events containing signals in the muon chambers were rejected and the 
visible energy (in the polar angle range between 20\gr and 160\gr) was required 
to be greater than 0.1$\sqrt{s}$. 
After the application of the preselection requirements, the data sample was 
dominated by Bhabha and Compton scattering events. 
At $\sqrt{s}$=206~GeV, they amounted respectively to about 81\% and 9\% of 
the preselected simulated sample, while the $e^+e^- \to \gamma \gamma (\gamma)$ events 
constituted about 3\%.
The remaining 7\% corresponded to $e^+e^-\to e^+e^-\gamma \gamma$ 
($\sim$4\%),
$e^+e^-\to \tau^+ \tau^- (\gamma)$ ($\sim$1\%) and $e^+e^-\to \nu\bar{\nu}\gamma \gamma$ ($\sim$0.7\%) events
and to residual contributions from other processes.
The next steps of the analysis consisted in first reconstructing photons, then selecting
multi-photonic final states and finally selecting $e^+e^- \to \gamma \gamma$ events.

\subsection{Photon reconstruction}

Photons are characterised by leaving energy in the electromagnetic calorimeters and 
no signals in the tracking devices, with the exception of photons converting
within the tracking system.
In both cases, the information provided by 
the calorimeters can be the input for photon reconstruction algorithms.
A clustering algorithm was therefore developed, consisting of a double cone 
centered on calorimeter energy deposits, in which the inner cone should contain the
energy deposited by the photon, while the absence of calorimetric deposits or
charged tracks beyond a predefined energy threshold in the outer cone should
ensure its isolation from other objects.

The parameters of the double-cone algorithm were optimized for the selection 
of the signal events using simulated \eeintogg(\fot) events. The best performance was
obtained by choosing two cones with vertex in the geometric centre of the
DELPHI detector and half-opening angles of 10\gr\ and 12\gr\ respectively. 
The energy in the inner cone was required to be above 5 GeV, while the total 
energy collected in the region between the two cones had to be below 5 GeV.

\subsection{Selection of multi-photonic events}

The main contamination to \eeintogg(\fot)\ events 
after preselection and photon reconstruction 
came from radiative Bhabha events 
with one non-reconstructed electron and the other electron lost in the beam pipe 
and from Compton scattering ($e^\pm\gamma$) events. 
Both backgrounds were dramatically reduced using the Vertex Detector as a veto 
against charged particles coming from the interaction point. 
Other sources of contamination of  
photonic final states such as calorimeter noise were eliminated by 
taking into account the profile of photonic showers in the electromagnetic and hadronic 
calorimeters. 
The photon identification criteria based on the response of the VD and of 
the calorimeters are listed below:

\begin{itemize}

\item
Charged particles coming from the interaction point were rejected by requiring that no 
VD track element\footnote{
A VD track element was defined by the presence of at least two signals in different  
VD layers separated in azimuthal angle
by at most 0.5\gr.} 
was within 3\gr\ in azimuthal 
angle\footnote{
The azimuthal angle, $\phi$, in the plane transverse to the beam direction, 
was defined with respect to the x-axis pointing to the centre of LEP.}
from photons reconstructed in the barrel. 
The corresponding angle in the endcaps was 10\gr.

\item 
In order to be compatible with the energy profile of photons in the HPC, 
particles reconstructed within the polar angle range $\theta \in [42^\circ,138^\circ]$ were 
required to either have energy deposits in at least 3 HPC layers  
- each containing more than 5\% of the electromagnetic energy of the particle -
or to have their azimuthal angle coinciding with the HPC modular 
divisions 
within $\pm$1$^\circ$.


\item 
Noise in the FEMC crystals was rejected by requiring that 
particles reconstructed within the FEMC acceptance 
had 
given signals in at most 50 lead-glass blocks and that   
at least 15\% of their electromagnetic energy was in one of the blocks.

\item 
Photons reaching the hadronic calorimeter should leave all their energy in 
its first layer, therefore, the reconstructed particles 
were required to have less than 5~GeV of hadronic energy associated to them 
or that at least 90\% of their hadronic energy component was recorded in the first 
layer of the HCAL.

\end{itemize}

\begin{figure}[H]
\begin{center}
\scalebox{0.4}{\includegraphics[bb=4 10 530 390]{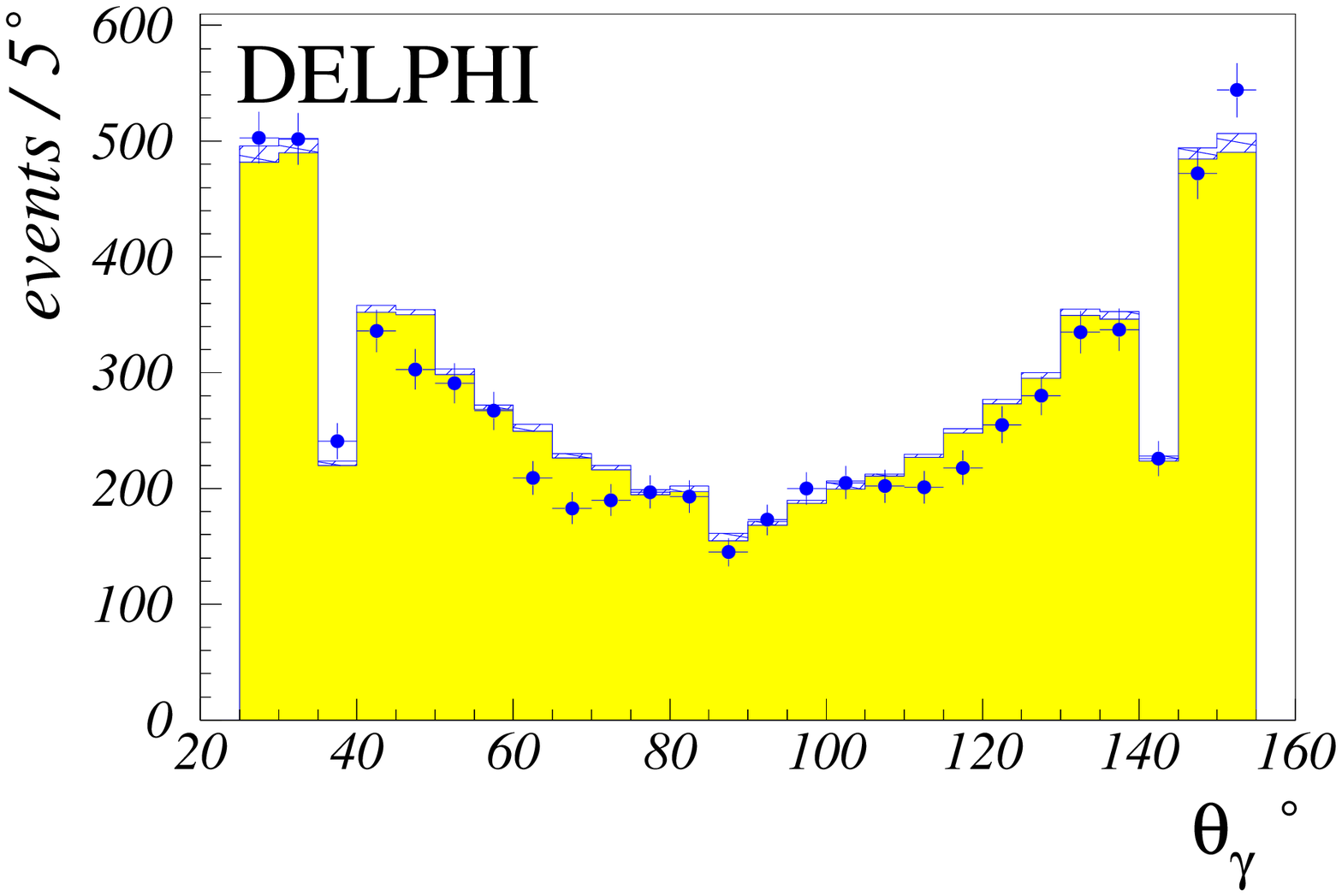}} 
\scalebox{0.4}{\includegraphics[bb=4 10 530 390]{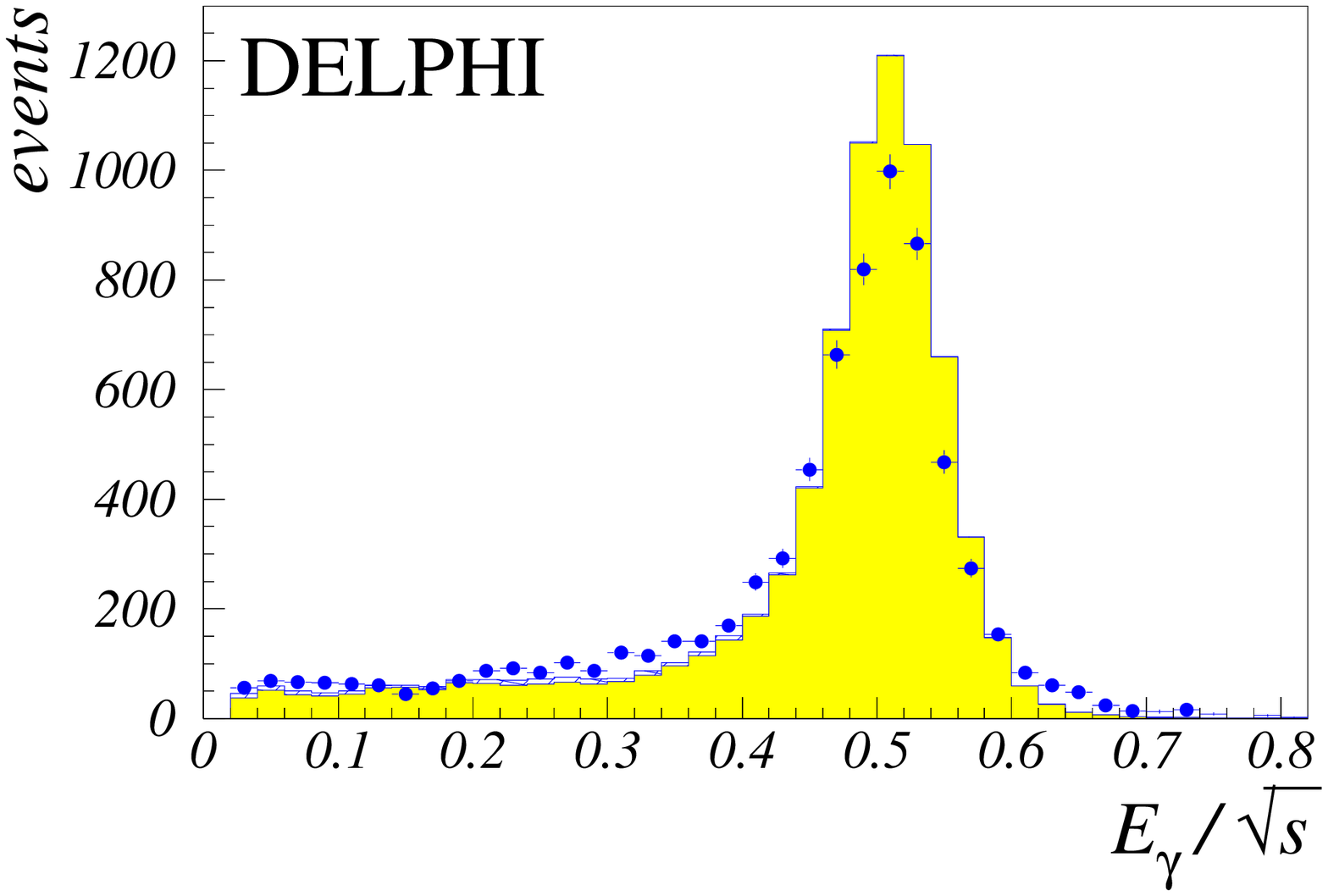}}\\
\scalebox{0.4}{\includegraphics[bb=4 10 530 390]{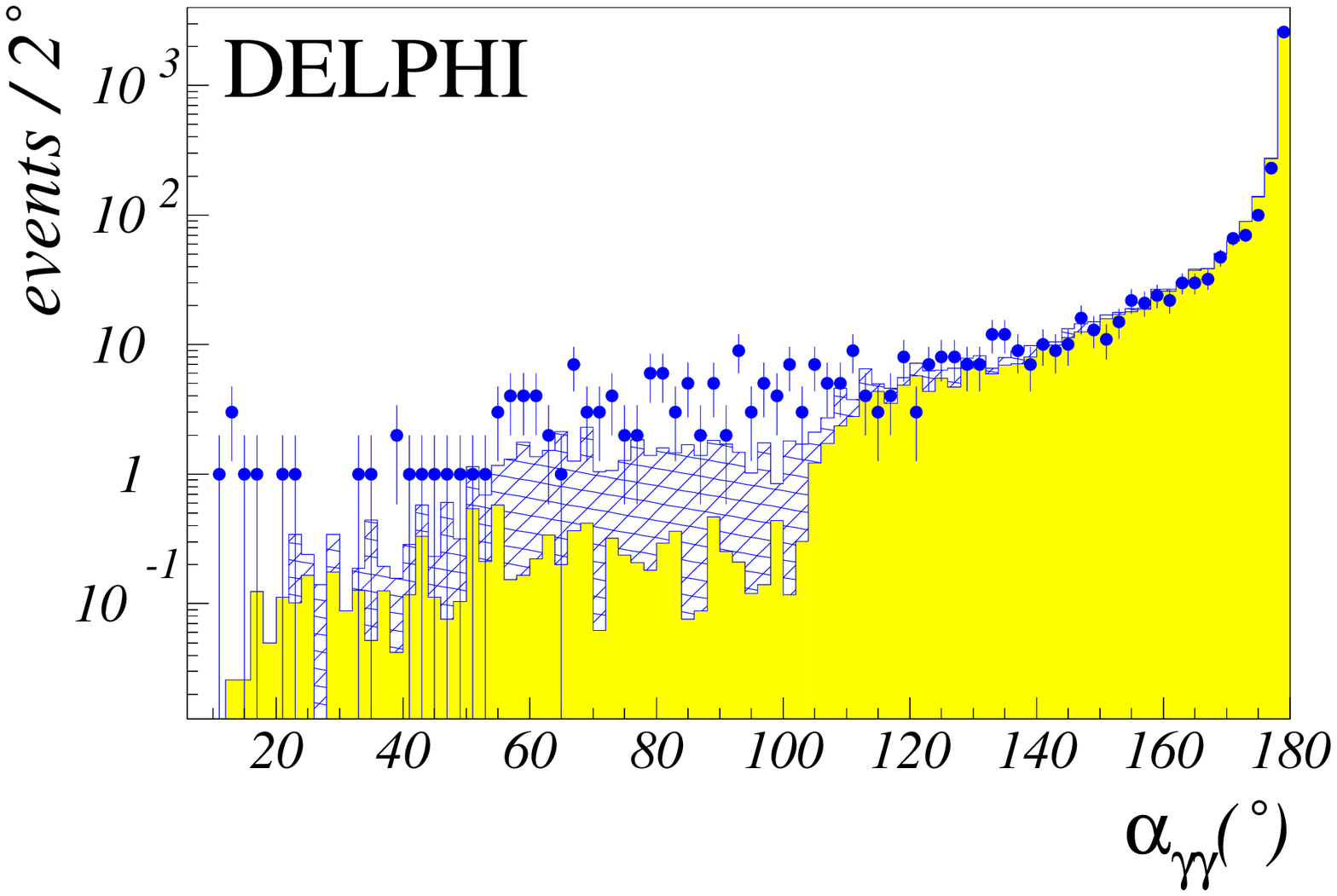}}  
\end{center}
\vspace*{-0.5cm}
\caption{
Distributions after the selection of multi-photonic events 
and before corrections for trigger and selection efficiency:
of the polar angle of the two most energetic photons (top left);
of their energy, normalised to $\sqrt{s}$ (top right) and of 
the angle between them (bottom) 
for the full data set (dots) compared to the SM predictions.
The shaded histograms correspond to the \eeintogg(\fot) expectations
while the hatched histograms correspond to the remaining processes:
Bhabha and Compton scattering and $\nu\bar{\nu}\gamma\gamma$ events.
 \label{fig:qedsel}}
\end{figure}

The selected sample consisted of events with 
at least two reconstructed particles 
fulfilling the photon reconstruction requirements and background rejection 
criteria described above. 
The two most energetic photons had to be located in the polar angle range between 25\gr\ and 155\gr\ and at most one converted photon, i.e. a reconstructed photon associated to well reconstructed 
charged particle tracks not associated to VD track elements, was allowed per event.
The numbers of events thus selected from real and simulated 
data are compared, as a function of $\sqrt{s}$, in table \ref{tab:multi_data-sim}.
Figure \ref{fig:qedsel} displays the selected event distributions of the polar angle, 
of the photon energy (normalised to $\sqrt{s}$) 
and of the angle between the two most energetic photons for the full data set 
and compares them to the Standard Model predictions.
For all distributions, the solid circles represent real data events, the shaded histogram 
represents the  $e^+e^-\to \gamma \gamma (\gamma)$ prediction and the hatched histogram 
the remaining background processes: 
Bhabha and Compton scattering and $e^+e^- \to \nu \bar{\nu} \gamma \gamma (\gamma)$ events.
An overall deficit of about 4.5\% was observed in the global data set with 
respect to the simulation predictions.
It is attributed to the photon trigger efficiency, which is lower in the 
barrel region of the detector, 
and to 
differences between the real detector and its simulated description, especially in 
aspects of the calorimeter performances.
Both effects are discussed and accounted for in section~\ref{sec:ggsec}.

\subsection{Selection of \beeintogg\ events\label{sec:ggsel}}

Events were considered as possibly due to the \eeintogg\ process if they 
contained two photons and no other reconstructed particles. 
Moreover, the two photons had to:

\begin{itemize}

\item 
have an energy greater than 0.15$\sqrt{s}$ each;

\item
be separated by a spatial angle of at least 130\gr;

\item
be contained within the polar angle acceptance of the VD 
and of the electromagnetic calorimeters: \\
$\theta \in  [25^\circ,35^\circ] \cup [42^\circ,88^\circ] \cup [92^\circ,138^\circ] \cup [145^\circ,155^\circ]$.

\end{itemize}


The \eeintogg\ selection resulted in a very low rate of expected background events.
The contamination from Bhabha events was already small after 
the selection criteria for multi-photon events were imposed (see table~\ref{tab:multi_data-sim}),
while the contribution from Compton scattering and $\nu\bar{\nu} \gamma\gamma$ events was drastically 
reduced mainly by the requirement that the spatial angle between both photons was large, but also by 
the energy cut.
The total contamination in the \eeintogg\ samples was found to be less than 0.2\%. 
Since this contamination corresponded to a very small number of background 
simulated events, 
it was fully taken into account as a systematic uncertainty in the determination 
of the cross-section for \eeintogg. The final numbers of selected and expected events are given in table \ref{tab:Nqed}.

\section{Determination of the Born level cross-section for the process \beeintogg(\bfot) \label{sec:ggsec}}

The analysis of the process \eeintogg\ was based on the study of 
distributions of $|\cos{\theta^\ast}|$ for the selected samples,
where $\theta^\ast$ is the polar angle of the photons with respect 
to the direction of the incident electron in the centre-of-mass of the $e^+e^-$ 
collision. $|\cos{\theta^\ast}|$ is defined by: 

\begin{equation}
|\cos \theta^\ast| = \sin [ 0.5 (\theta_{\gamma 1}-\theta_{\gamma 2}) ] /
\sin [0.5 (\theta_{\gamma 1}+\theta_{\gamma 2})] 
\label{eq:thstar}
\end{equation}

\noindent 
where $\theta_{\gamma 1}$ and $\theta_{\gamma 2}$ are the polar 
angles of the photons in the laboratory frame.
This parameterisation of the polar angle of the photons 
enables the measurement of the differential cross-section of \eeintogg\ 
to be insensitive to Initial State Radiation photons radiated collinearly to 
the beam.

In the present analysis, 
the retained $|\cos\theta^\ast|$ acceptance was divided into 16 intervals:
the barrel part of the detector, corresponding to 
$[0.035,0.731]$ with 14 bins (each covering $|\Delta\cos\theta^\ast|$ = 0.0505,
except for the last two, for which  $|\Delta\cos\theta^\ast|$ = 0.045)
and the forward region, corresponding to $[0.819,0.906]$, which was divided  
in two equal bins.

\subsection{Evaluation of the selection efficiency\label{sec:effsel}}

The efficiency for selecting the $\gamma \gamma$ samples was evaluated as a function of 
$|\cos{\theta}^\ast|$ using simulated sets of $e^+e^- \to \gamma \gamma(\gamma)$ events 
generated at different centre-of-mass energies. The average selection efficiency
in the barrel region of DELPHI was close to 80\%, whereas in the endcaps
it was found to be about 52\% for data taken prior 
to 1999 and $\sim$62\% for the remaining data, resulting from 
improvements in the particle reconstruction algorithms used for the forward region of DELPHI.

Small differences between the real detector response  and the detector simulation were observed.
These differences came from aspects of the performance of the calorimeters, 
and were essentially  caused  by the limited accuracy of the simulation 
description in the edges of HPC modules.
Differences due to the description of the amount of material in 
front of the calorimeters 
could also arise, especially in the endcaps.
In order to investigate these differences, sets of real and simulated Bhabha 
events were selected using the information provided by the 
tracking devices alone.
The two photon selection criteria based on calorimeter information 
were then imposed on the selected samples.
The differences between the effect of the selection criteria  
on real and simulated data were then parameterised as a correction factor ${\cal R}$,
which was taken as the ratio between simulated and real data rates of 
Bhabha events fulfilling all criteria.
The correction factor ${\cal R}$ reflects the necessary correction to the 
selection efficiency as evaluated from the simulation of \eeintogg(\fot).


The corrections to the selection efficiency were 
evaluated as a function of the $|\cos{\theta^\ast}|$ interval
and of $\sqrt{s}$. Their average value for each analysed data set 
was found to be above 1.0 by 1\% to about 2.8\%, depending on the data set, 
reflecting an over-estimation of the selection efficiency. The main contribution for this difference was found to come from 
the barrel region of DELPHI, with exception for the 206.3~GeV data set where 
the correction to the selection efficiency in the endcaps amounted to 4.7\%.
The average selection efficiencies, for the barrel and endcaps of DELPHI,
including the corresponding corrections are displayed in table \ref{tab:Nqed}
as a function of $\sqrt{s}$. 

\subsection{Evaluation of the trigger efficiency \label{sec:trigger}}

The trigger efficiency for $\gamma \gamma$ final states without 
photon conversions was 
computed with Bhabha ($e^+e^- \to e^+e^-$) events and using the 
redundancy of the electromagnetic trigger with the track trigger. 
It was evaluated as a function of $|\cos{\theta^\ast}|$
for each data set and taking into account the different trigger settings.


The $\gamma \gamma$ trigger efficiency values in the forward region were close to 100\% 
for all data sets.
As for the barrel region, average values close to 97\% were obtained.
The lower value obtained for the 172~GeV data set (78\%) 
was due to a malfunctioning of the barrel single photon trigger, present during 
part of the 1997 data taking period.
The average trigger efficiencies for the barrel and endcaps are displayed 
in table \ref{tab:Nqed}.


Final states with one converted photon, 
were triggered by the single track coincidence trigger, 
whose efficiency was nearly 100\% \cite{bib:trigger}.

\subsection{Radiative corrections}

The event selection does not correspond to the
Born level contribution from $e^+e^- \to \gamma \gamma$ 
since no selection criteria can remove 
the higher order contributions coming from
soft {\it bremsstrahlung} and the exchange of virtual gauge bosons. 
In order to take such higher order corrections into account, the radiative effects were 
estimated using the \eeintogg(\fot)\ physics generator of Berends and Kleiss \cite{bib:BerKl} 
which computes the cross-section for \eeintogg\ up to ${\cal{O}}(\alpha^3)$.
The radiative correction factor $R_i$ was evaluated  as a function of $|\cos{\theta^\ast}|$
by generating 
$10^7$ events at the average centre-of-mass energy value corresponding to each data set, although 
its variation within the considered $\sqrt{s}$ range was negligible.
It was taken as the ratio between the \eeintogg(\fot)\ cross-section 
computed up to order $\alpha^3$ to the Born level cross-section (${\cal{O}}(\alpha^2)$),
which for a given $\Delta\cos\theta^\ast$ acceptance 
(including the  $\theta^\ast$ complement with respect to 90\gr) is given by:

\begin{equation}
\sigma^0_{QED}(\Delta\cos\theta^\ast) = \frac{2\pi \alpha^2}{s} \biggl[
\ln{\frac{1+\cos{\theta^\ast}}{1-\cos{\theta^\ast}}}- \cos{\theta^\ast} \biggr]_{\Delta\cos{\theta^\ast}}
\label{eq:stotQED}
\end{equation}

The radiative correction factor ranged between 1.03 for events with 
high $|\cos{\theta^\ast}|$ values (forward events) and about 1.07 
for low values of $|\cos{\theta^\ast}|$, and are displayed as a function of 
$|\cos{\theta^\ast}|$ in table~\ref{tab:radcor}.

\subsection{Systematic uncertainties}

The systematic uncertainty on the cross-section measurement 
was obtained for each $|\cos\theta^\ast|$ interval and for the ten data subsets by 
adding in quadrature the respective relative  residual background expectations,
the uncertainties on the selection efficiency
and on the corresponding correction, the uncertainty on the trigger determination, 
the statistical error on the radiative corrections determination and 
the uncertainty from the determination of the luminosity.

The uncertainty on the determination of the selection efficiency 
had three components, 
coming from the statistical uncertainty due to the finite Monte Carlo statistics, 
from the uncertainty on its correction 
and from the experimental resolutions of the variables used in the \eeintogg\ selection.
The latter were, as described in sections~\ref{sec:ggsel} and~\ref{sec:ggsec}, 
the angle between the two photons, their energy and $|\cos{\theta^\ast}|$.
The uncertainties were evaluated, 
for each of the \eeintogg(\fot) simulated samples, 
from the distributions of the difference between the values of the 
corresponding variables at generator level and 
after being passed through the full detector simulation and reconstruction chain.
The uncertainties on the energy of the photons were taken as the r.m.s. values of  
Gaussian fits to these difference distributions.
The same procedure was followed for the uncertainty on $\cos\theta^\ast$ which was 
evaluated independently for each $|\cos\theta^\ast|$ interval. 
The uncertainty on the spatial 
angle between the photons was taken as the quadratic sum of the shift in the 
central value and of half of the full width at half maximum of the distributions.

The uncertainty in the luminosity determination had both experimental and theoretical 
contributions. The experimental contribution corresponds to a
$\pm$0.5\% systematic uncertainty in the measurement of the luminosity.
The theoretical contribution to the luminosity determination uncertainty 
was taken to be $\pm$0.25\%, from the Bhabha event generator BHLUMI~\cite{bib:BHLUMI}.
A total error of $\pm$0.56\% on the luminosity determination was therefore obtained 
by adding in quadrature both the experimental and theoretical uncertainties. 

The overall values for the systematic uncertainty associated to each data sample 
are presented in table~\ref{tab:esyst}, as well as the specific contributions 
coming from different sources of systematic error.
The dominant contribution to the systematic error came from the 
uncertainty on the determination of the selection efficiency.

\subsection{Cross-section for \beeintogg }

The differential cross-section for \eeintogg\ was computed for each 
$|\cos{\theta^\ast}|$ interval~($i$) as:  

\begin{equation}
\frac{d\sigma^0_i}{d\Omega}= \frac{1}{2\pi (\Delta \cos{\theta^\ast})_i}
\frac{1}{\cal{L}}\frac{N^{\gamma\gamma}_i}{\varepsilon_i R_i}
\label{eq:sdifd}
\end{equation}

\noindent
where ${\cal{L}}$ is the integrated luminosity of the considered data set,
$N^{\gamma\gamma}_i$ is the corresponding  number of selected events, 
$\varepsilon_i$ is the corresponding product 
of trigger efficiency and selection efficiencies 
(including corrections) and $R_i$ is the radiative correction factor.
The numbers of events selected from each data sample, 
are compared to the simulation predictions (accounting for trigger efficiency
and for selection efficiency corrections) in table~\ref{tab:Nqed}.
A total number of 2679 events were selected from data while 
2761$\pm$15 were expected from the \eeintogg(\fot) 
simulation, taking into account the trigger efficiency and 
the selection efficiency corrections. 
In table \ref{tab:nev}, the number of events and efficiency 
(the product of trigger and selection efficiencies) 
are displayed for each data set as a function of $|\cos{\theta^\ast}|$. 

For the purpose of combining the 
DELPHI data with the data taken by the other LEP experiments, 
a different $|\cos{\theta^\ast}|$ binning is used in table \ref{tab:lwg} 
to display the numbers of events and the analysis efficiency 
for a subsample of 2206 events, among the the 2679 events selected.
The corresponding radiative correction factors are displayed as a function of $|\cos{\theta^\ast}|$ 
in table~\ref{tab:radcor}.

\begin{figure}[H]
\noindent
\scalebox{0.4}{\includegraphics[bb=10 10 530 320]{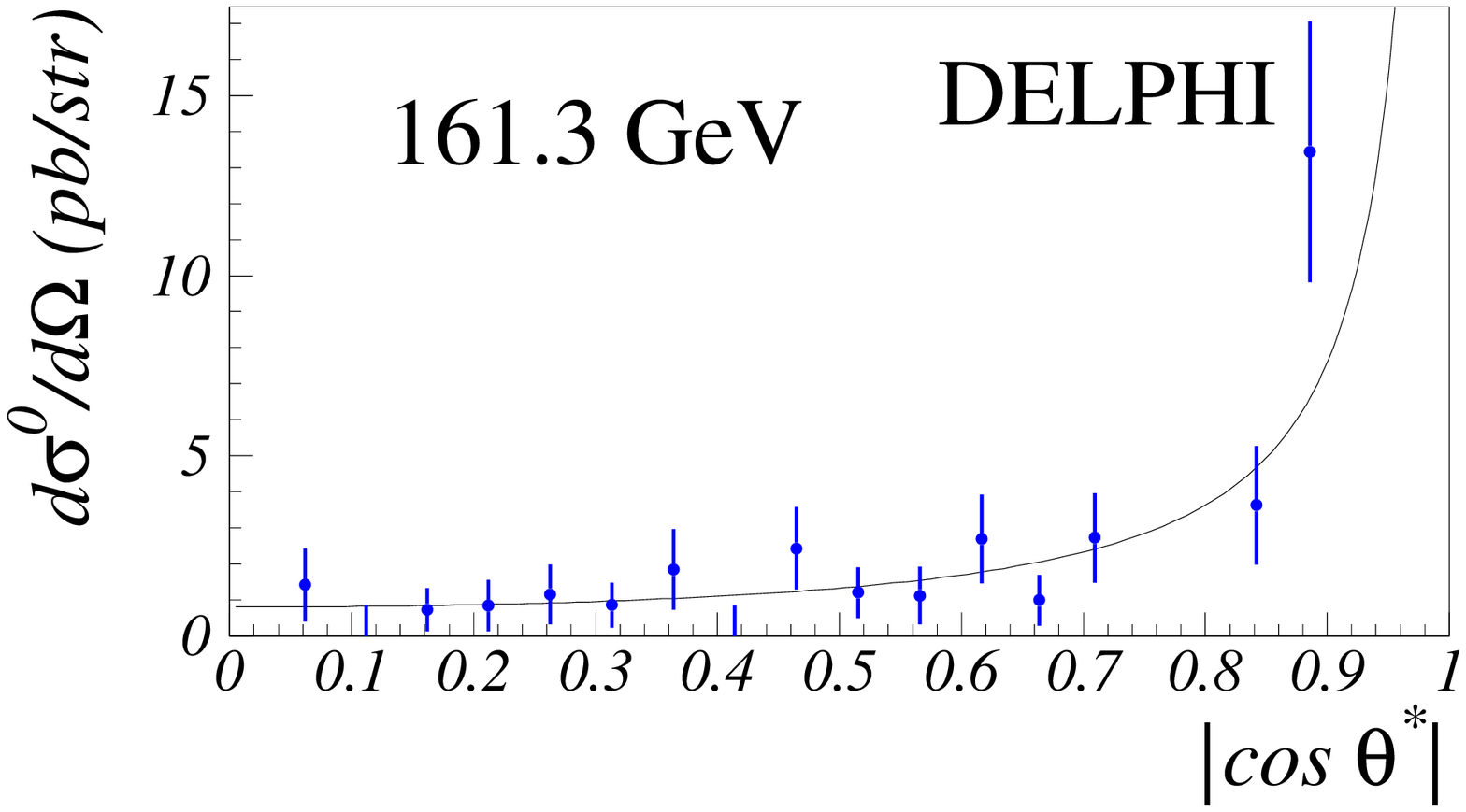}}
\scalebox{0.4}{\includegraphics[bb=10 10 530 320]{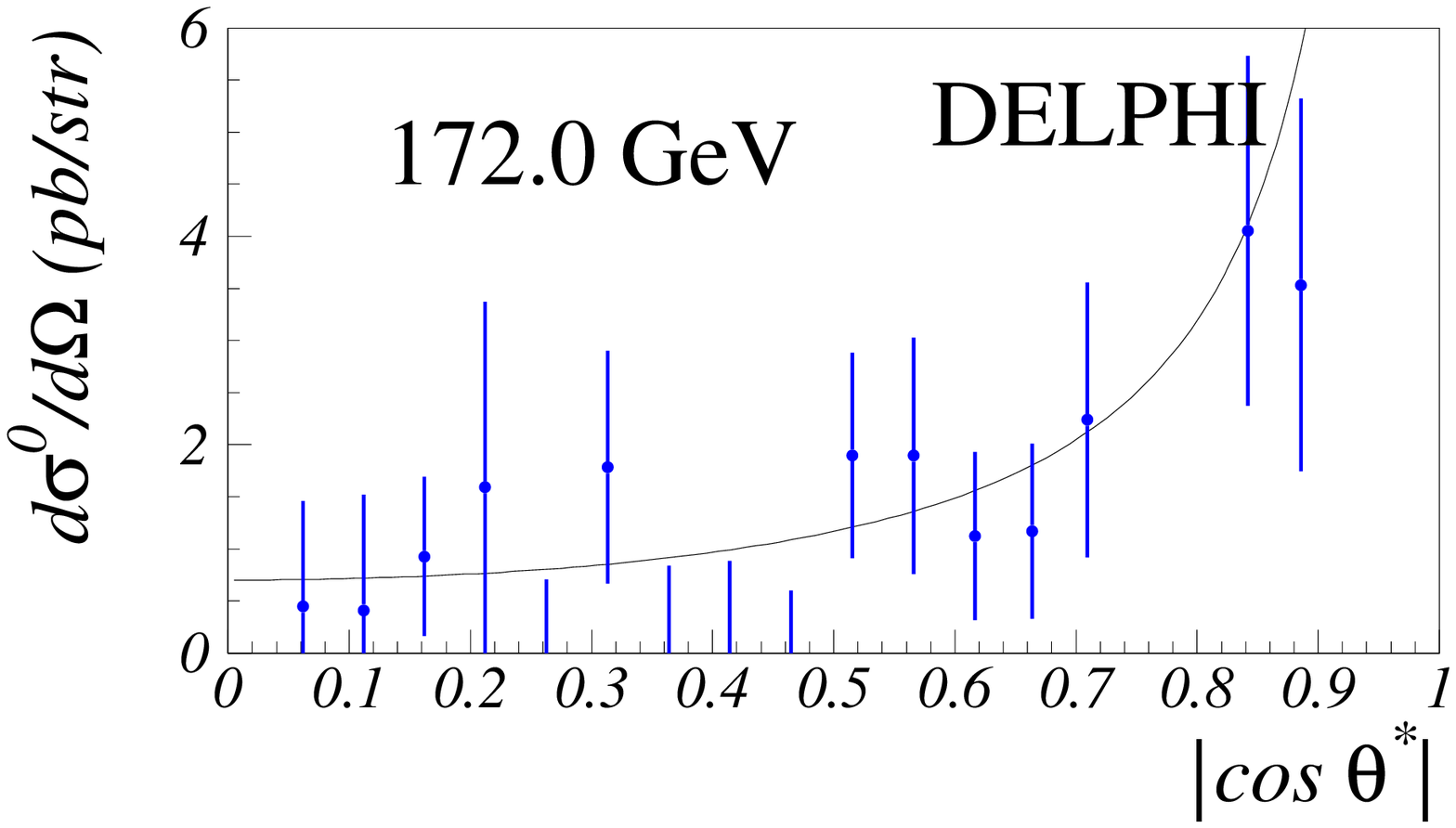}} \\
\scalebox{0.4}{\includegraphics[bb=10 10 530 320]{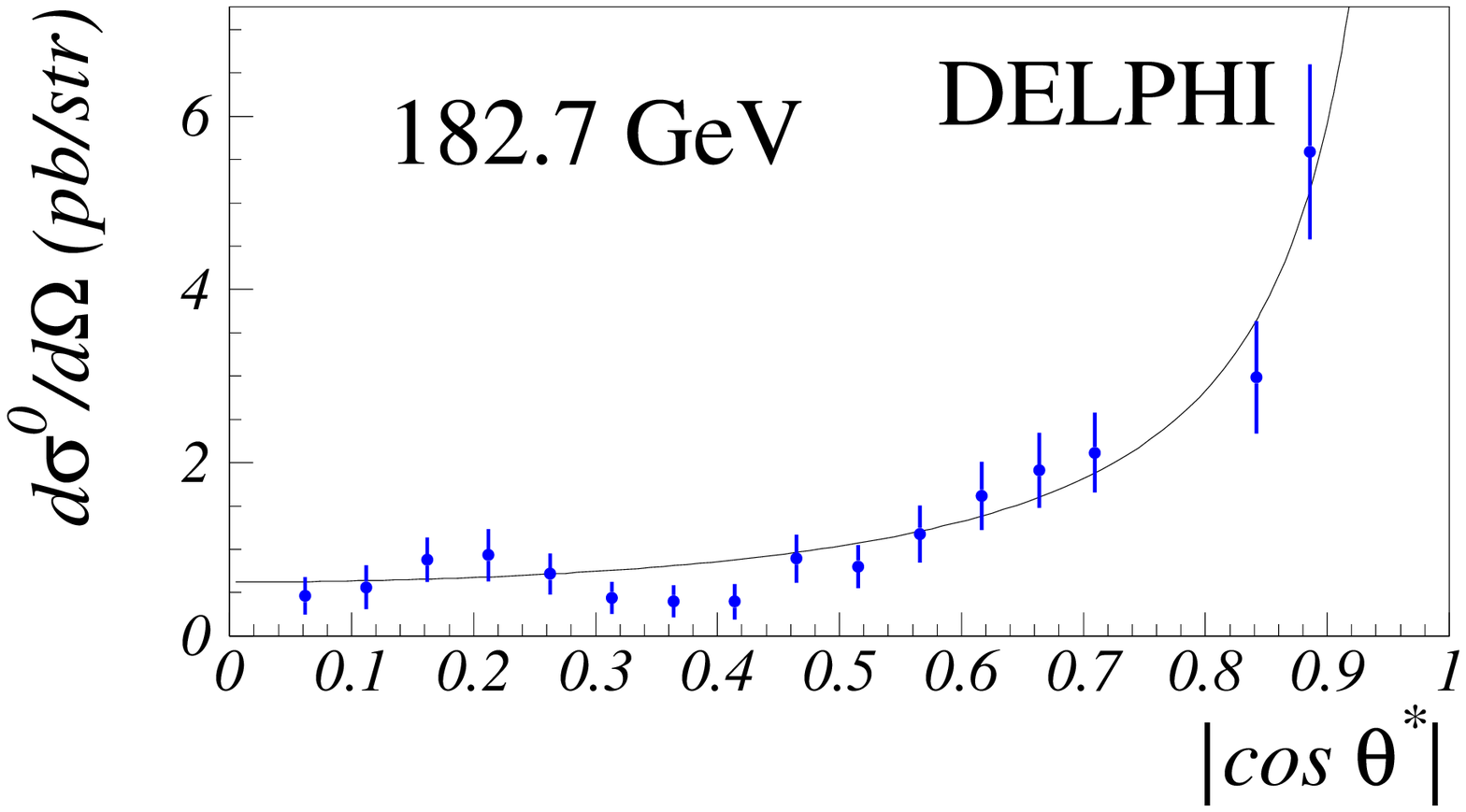}}   
\scalebox{0.4}{\includegraphics[bb=10 10 530 320]{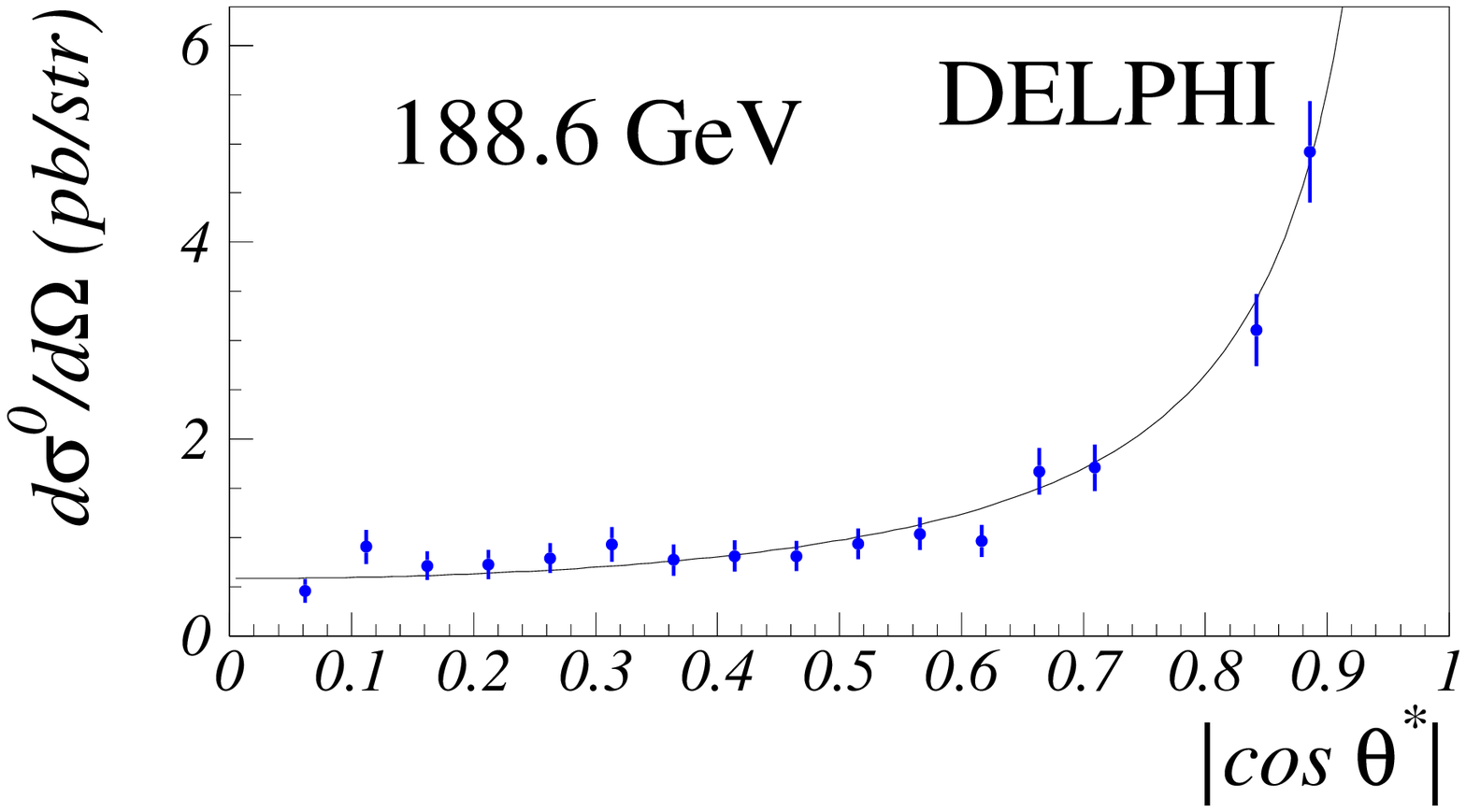}} \\
\scalebox{0.4}{\includegraphics[bb=10 10 530 320]{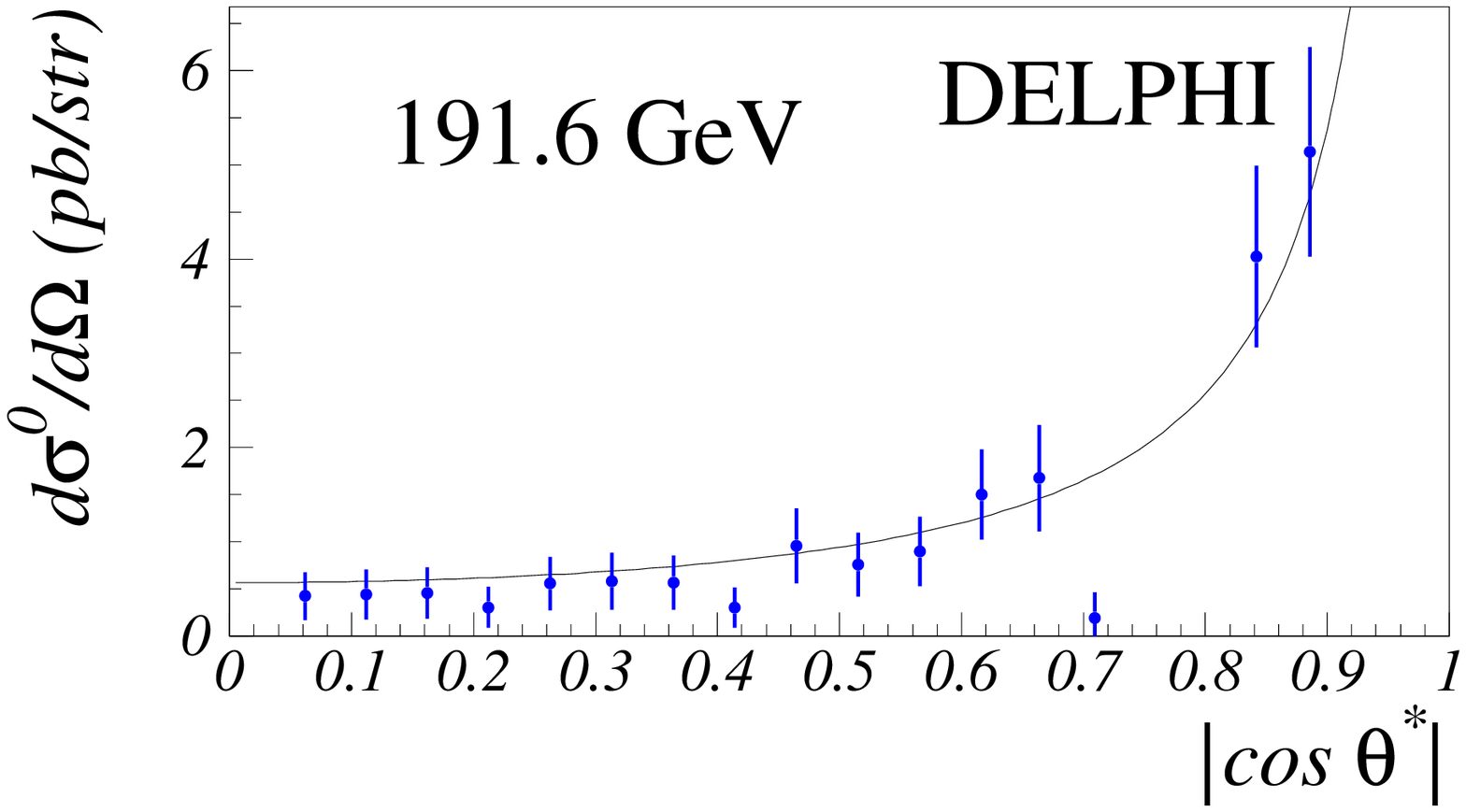}} 
\scalebox{0.4}{\includegraphics[bb=10 10 530 320]{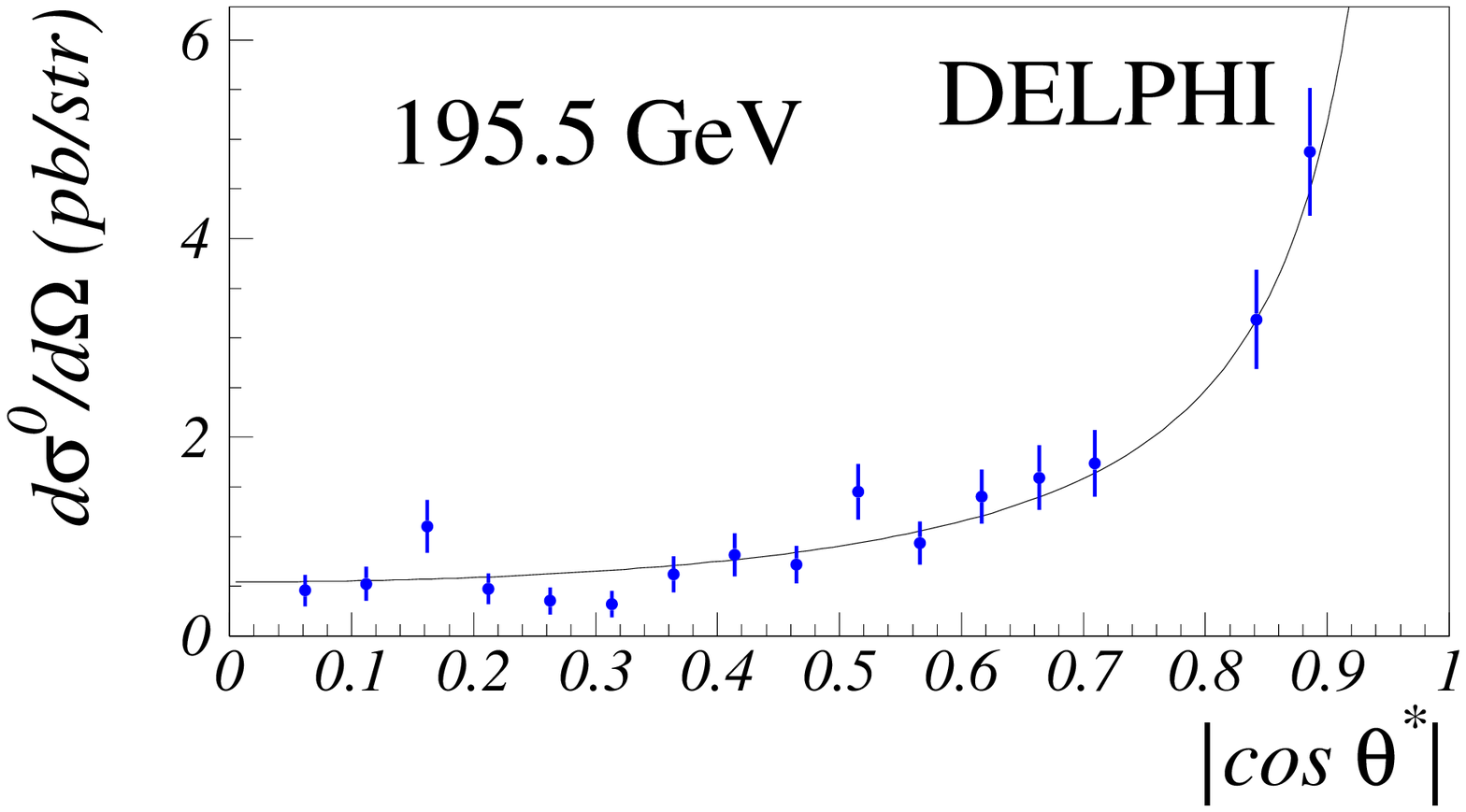}}\\
\scalebox{0.4}{\includegraphics[bb=10 10 530 320]{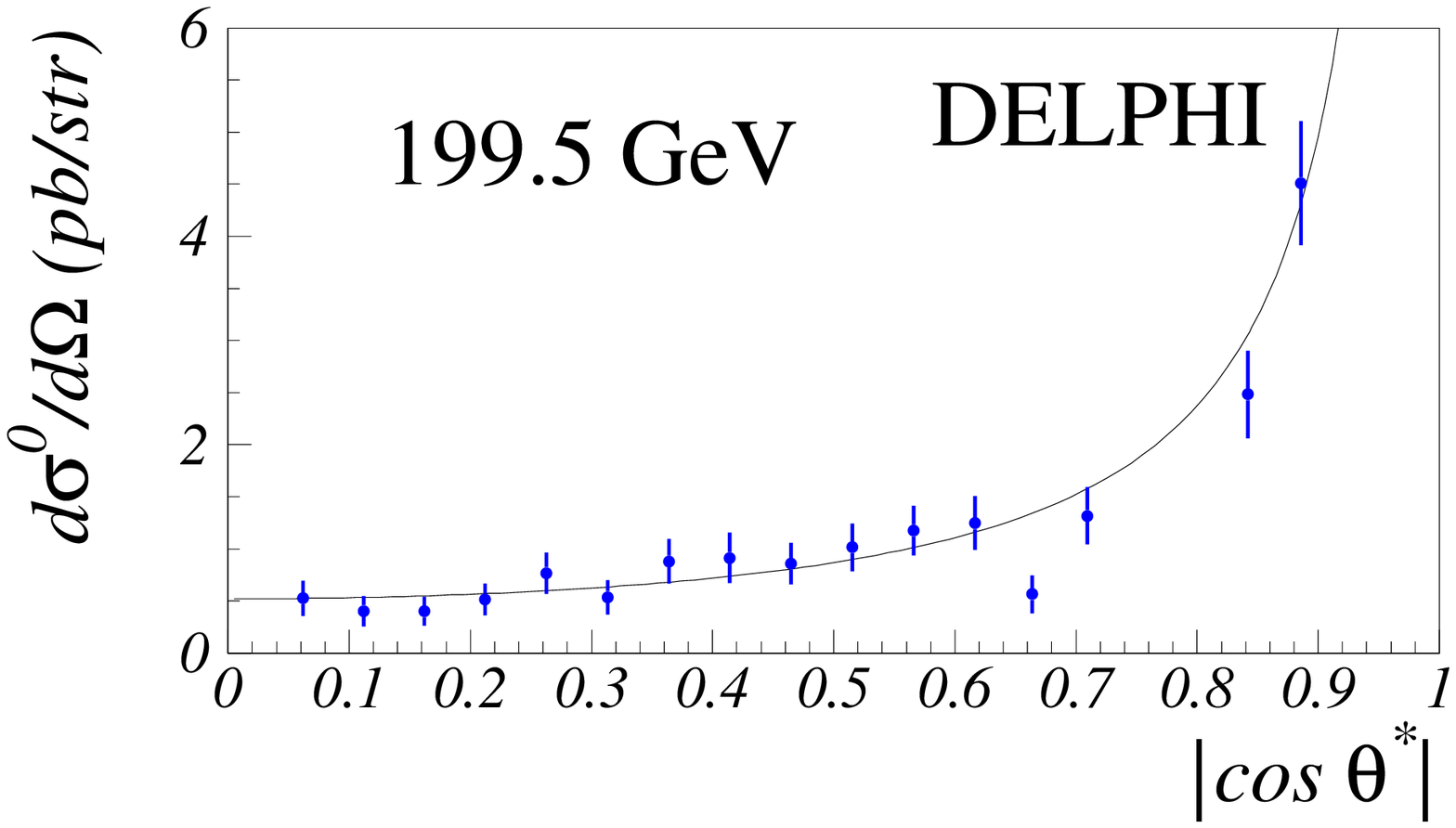}} 
\scalebox{0.4}{\includegraphics[bb=10 10 530 320]{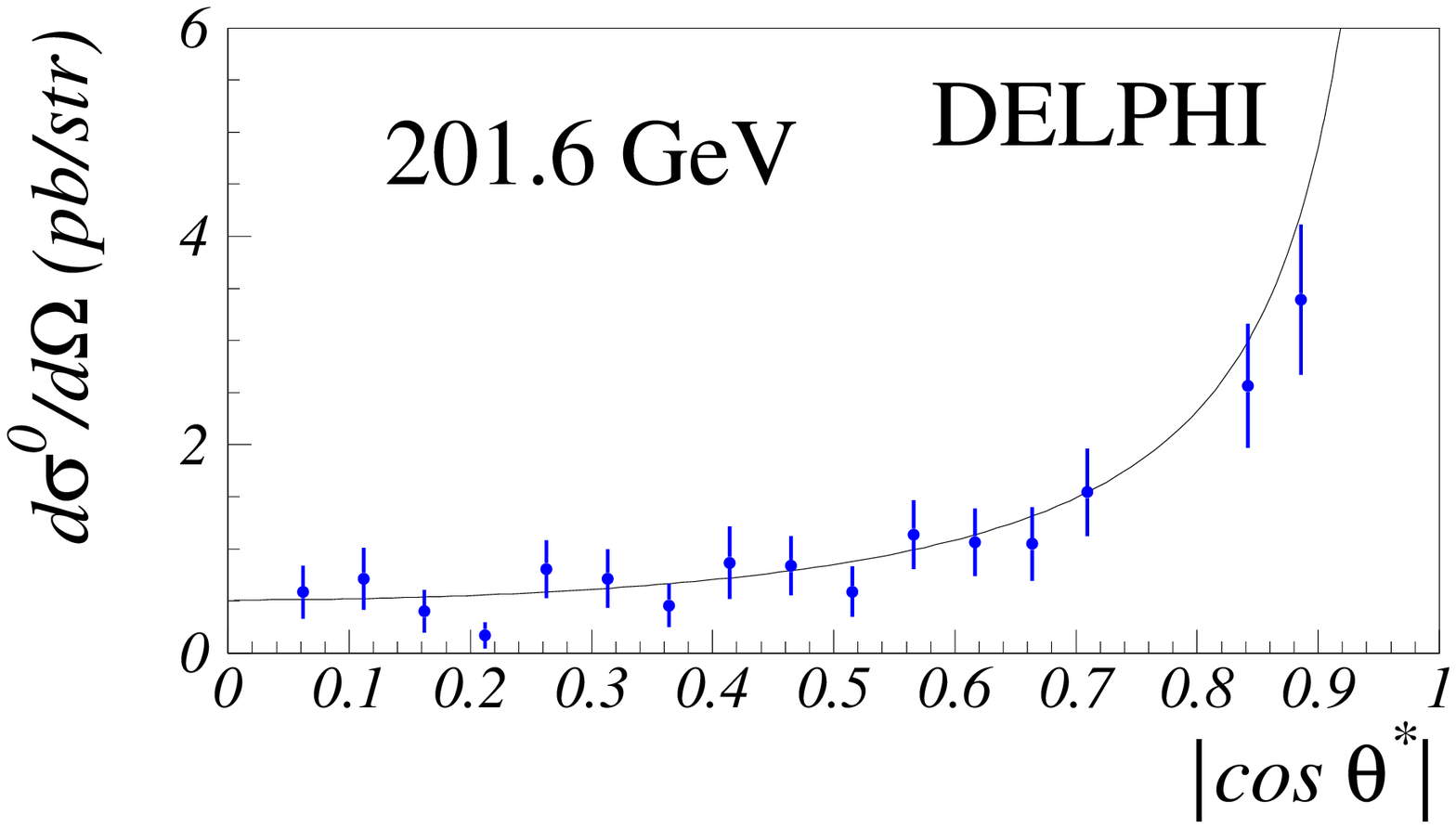}}\\
\scalebox{0.4}{\includegraphics[bb=10 10 530 320]{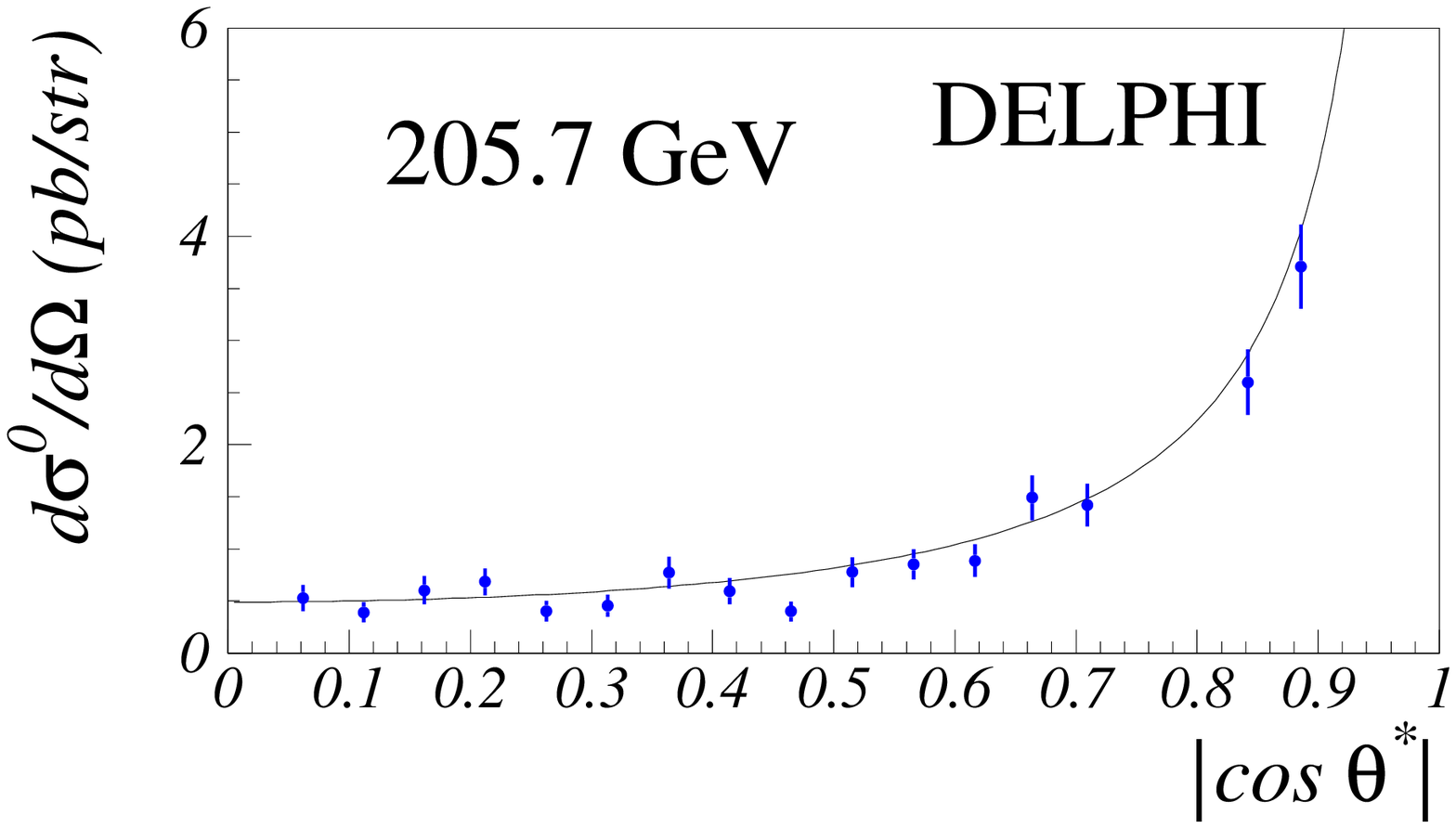}}
\scalebox{0.4}{\includegraphics[bb=10 10 530 320]{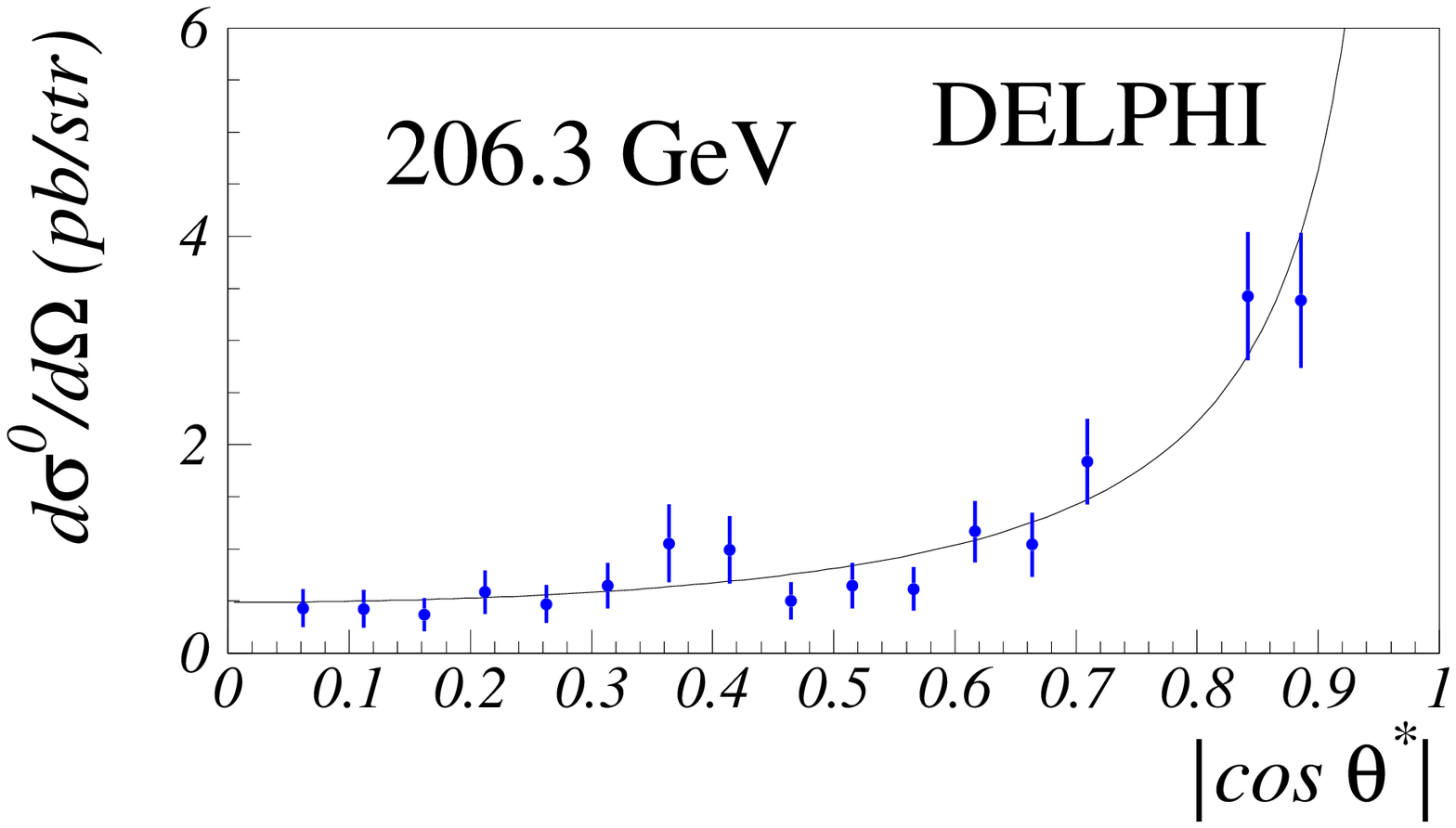}} 
\vspace*{-0.5cm}
\caption{Differential Born level cross-sections obtained from the 
ten data sets, compared to the corresponding QED predictions.
The error bars represent the statistical and systematic errors 
added in quadrature. 
Note the different scale used for displaying the 161.3~GeV data.
\label{fig:dsdt_i}}
\end{figure}

\begin{figure}[H]
\begin{center}
\scalebox{0.6}{\includegraphics[bb=13 27 524 430]{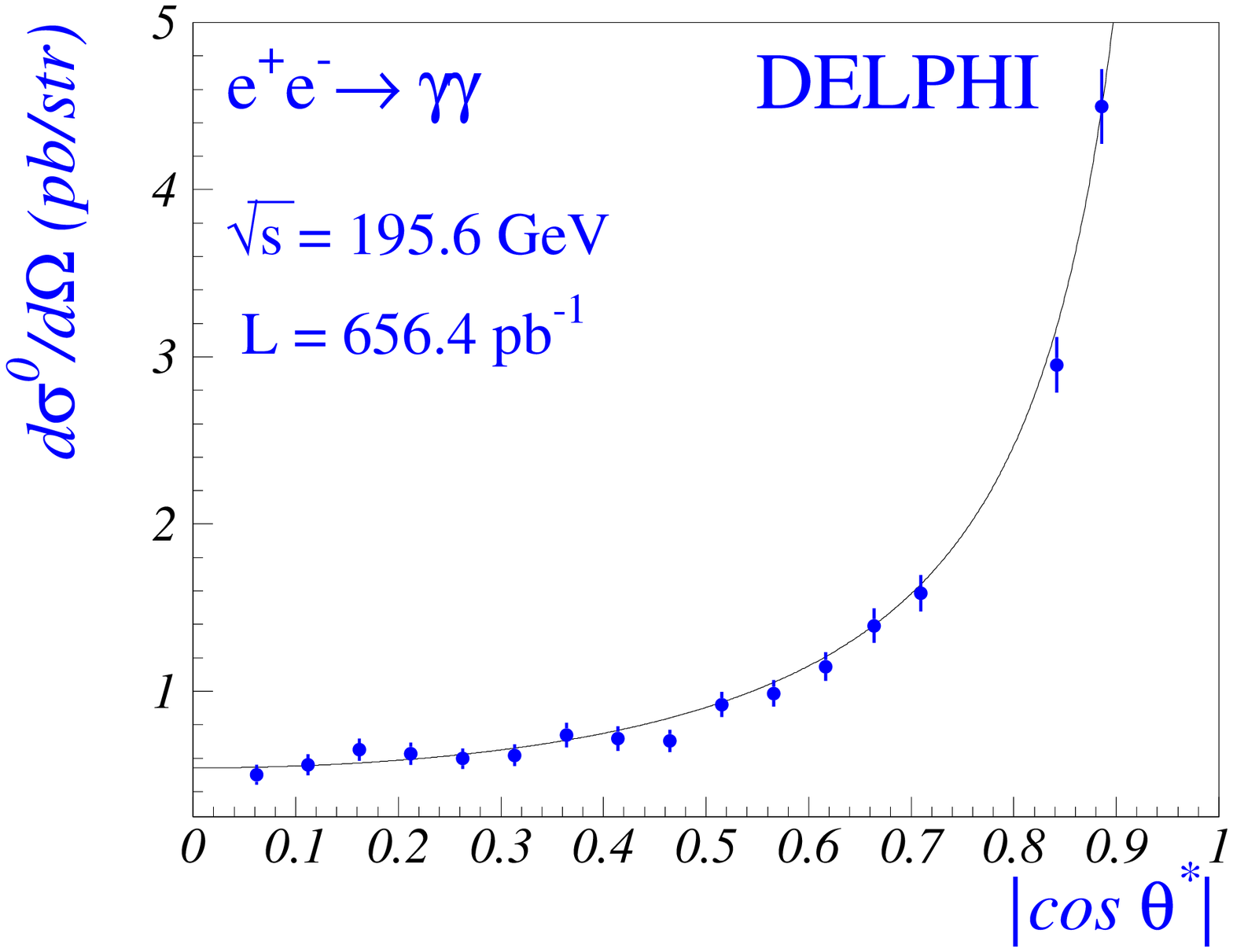}} \\
\scalebox{0.6}{\includegraphics[bb=13 23 524 320]{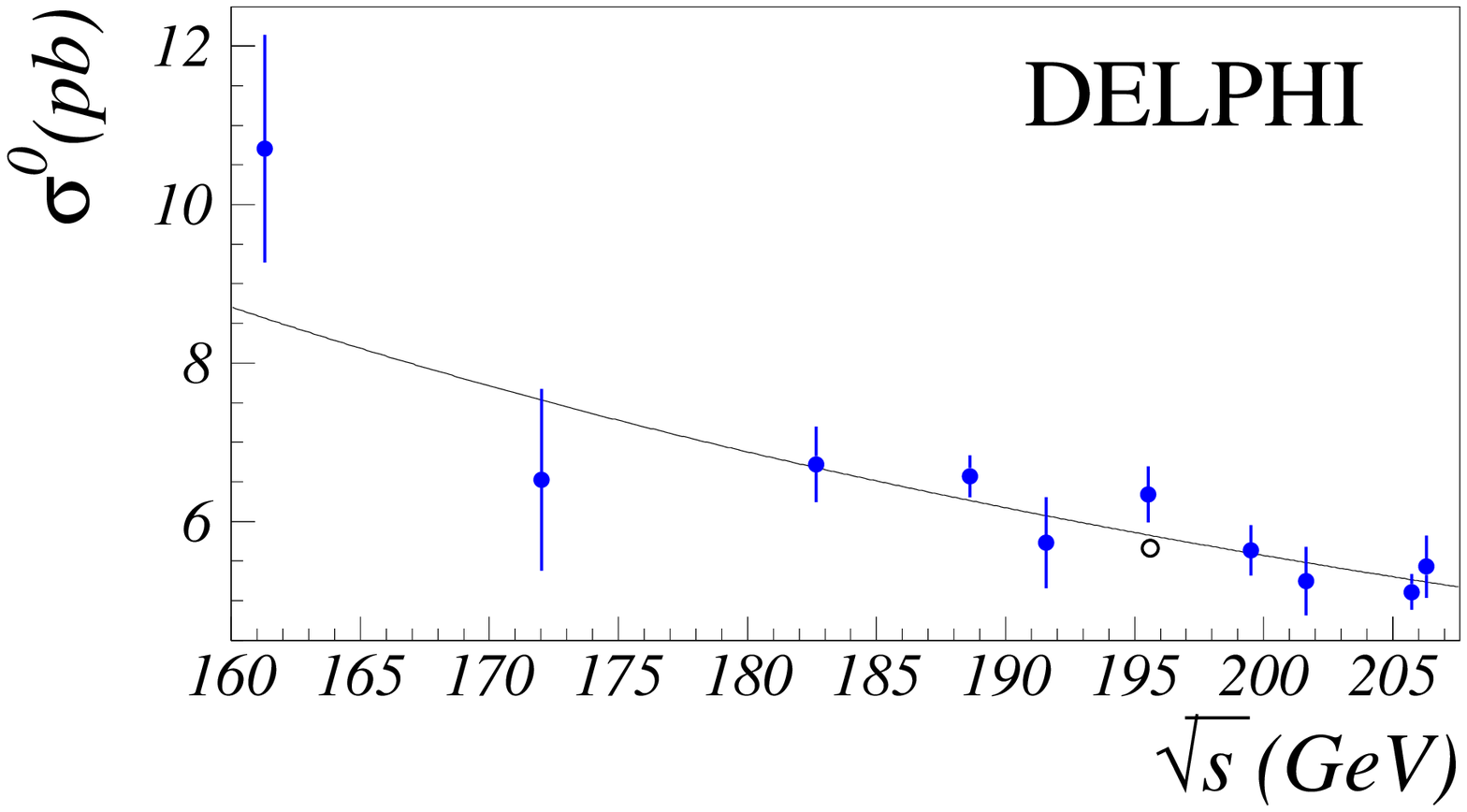}} \\
\end{center}
\vspace*{-0.5cm}
\caption{
Top: The Born level differential cross-section for \eeintogg(\fot) obtained by combining the ten 
data sets at an effective 
centre-of-mass energy of 195.6~GeV (dots) compared to the QED theoretical distribution 
(full line). 
Bottom: The visible Born level cross-section for each of the ten data sets (dots) 
as a function of the centre-of-mass energy. The empty circle corresponds to the 
average visible Born cross-section for all LEP 2 data: 5.66$\pm$0.11$\pm$0.03~pb.
The errors have been estimated by adding in quadrature the 
statistical and the systematic uncertainties associated to  the measurements. 
\label{fig:sigma}}
\end{figure}

The Born level differential cross-section distributions for the ten data sets 
are compared in figure~\ref{fig:dsdt_i} to the corresponding theory predictions.
An average centre-of-mass energy of~195.6~GeV was calculated 
for the entire data sample by weighting the integrated luminosities 
of the ten data sets by the corresponding $1/s$ factor, 
in order to take into account the dependence of the Born level cross-section on the 
centre-of-mass energy (equation~\ref{eq:stotQED}).
The average differential cross-section is compared to the theoretical prediction in 
figure~\ref{fig:sigma}~(top) and in table~\ref{tab:dsdo}. 
It was computed for each $|\cos\theta^\ast|$ interval from expression~\ref{eq:sdifd}, 
where $\cal L$ was the integrated luminosity of the full data set
and the third fraction was replaced by the sum of the number of events in each bin  
divided by the corresponding correction, 
consisting of the product of the efficiency and of the radiative correction factor.

The total visible cross-section for the process \eeintogg, integrated over the  
full $|\cos{\theta^\ast}|$ acceptance, i.e. $|\cos\theta^\ast|\in[0035,0.731] \cup [0.819,0.906]$,
was evaluated for each centre-of-mass energy, using the expression:   

\begin{equation}
\sigma^0_{dat}= \frac{1}{\cal L} \sum_{i=1}^{Nbins} \frac{N^{\gamma\gamma}_i}{\varepsilon_i R_i}
\label{eq:stot}
\end{equation}

The visible cross-section measurements as a function of $\sqrt{s}$  
are compared to theory in figure~\ref{fig:sigma}~(bottom) 
and their values 
(with statistical and systematic uncertainties) are displayed in table~\ref{tab:stotqed}.
The $\chi^2/ndof$ of the measured differential cross-section distributions 
with respect to theory\footnote{The theoretical uncertainty on the QED prediction, estimated to be below $\pm$1\%, was neglected.}
were computed for each data set and are also 
presented in table~\ref{tab:stotqed}, showing the agreement between the measurement and theory.  
The total visible cross-section at the average centre-of-mass energy of 195.6~GeV 
was computed as the ratio between the sum of the number of events found in each 
bin and for all data sets, 
corrected by the corresponding $1/(\varepsilon_i R_i)$ factors, 
and the integrated luminosity of the full data sample. The obtained cross-section was 
$5.66\pm0.11\pm0.03$~pb, in agreement with the $5.83$~pb expectation from the 
Standard Model.

\begin{figure}[H]
\begin{center}
\scalebox{0.6}{\includegraphics[bb=13 23 524 320]{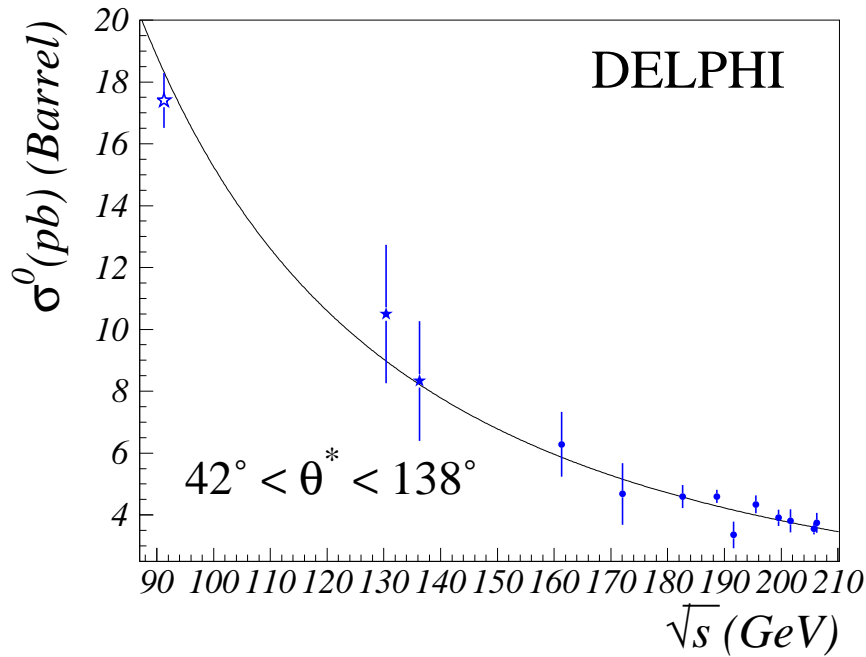}}
\end{center}
\vspace*{-0.8cm}
\caption{Born level cross-section for \eeintogg\ 
in the barrel region of DELPHI,
$42^{\circ}<\theta^{\ast}<138^{\circ}$, as a function of the 
centre-of-mass energy, 
for 1990-1992 LEP 1 data (white star),
LEP 1.5 data collected in 1995 and 1997 (black stars),
and for LEP 2 data collected between 1996 and 2000 (dots), 
compared to the QED prediction.
The errors have been estimated by adding in quadrature the 
statistical and the systematic uncertainties associated to  the measurements. 
\label{fig:sigma_lep2}}
\end{figure}

The Born level cross-section measurements in
the region $0.035 < |\cos{\theta^\ast}| < 0.731$, 
were corrected to the full barrel acceptance of DELPHI, 
$0.000 < |\cos{\theta^\ast}| < 0.742$, 
and the obtained values are displayed in figure \ref{fig:sigma_lep2}
as a function of the centre-of-mass energy, 
along with previously published results, derived from 
LEP 1 data collected between 1990 and 1992~\cite{bib:94} 
and from LEP 1.5 data collected at $\sqrt{s}$=130-136~GeV~\cite{bib:96-97}.

\section{Deviations from QED}

Possible deviations from QED are described in the context of several different models 
in which the Born level differential cross-section for \eeintogg\ is expressed, 
as in (\ref{eq:sigdev}), as the sum of the QED term and of a deviation term, the latter 
parameterised as a function of an energy (or mass) scale of relevance for the model tested.

\begin{equation}
\frac{d\sigma^0_{i}}{d\Omega}=
\frac{\alpha^2}{s}\biggl(
\frac{1+\scos \theta^\ast}{1-\scos \theta^\ast} \biggr)+ 
\biggl (\frac {d\sigma}{d\Omega}\biggr)^D
\label{eq:sigdev}
\end{equation}

The 95\%~C.L. lower limits on the free parameters \--- energy/ mass scales \--- 
for different models were extracted using binned maximum likelihood functions.
These were built as the joint 
probabilities for the number of observed events in each $|\cos \theta^\ast|$ bin 
and data set, given the number of expected events from theory.
For each model, the number of expected events was a function of 
the corresponding parameter, written in the form of an estimator $\xi$, 
which, whenever possible, was chosen to yield an approximately Gaussian 
distribution for the likelihood function.
The systematic uncertainties were taken into account as the r.m.s. values of 
Gaussian probability density functions of free normalization parameters.
The maximization of the likelihood functions was performed using 
the program MINUIT~\cite{bib:MINUIT}. 
For each model, the results were translated in terms of central $\bar{\xi}$ 
and positive and negative r.m.s. values $\sigma_\pm$ for the 
estimators, whereas the 95\%~C.L. lower limits 
for the parameters were obtained by renormalising the 
joint probability distribution to their physically allowed region~\cite{bib:pdg}. 

The different models and the 95\%~C.L. lower bounds derived for the 
corresponding parameters presented in detail in the following sections are 
summarised in table~\ref{tab:devQED}.

\subsection{QED cutoff}

The most general way to parameterise a deviation from QED is by the introduction of 
a QED cutoff, representing the energy scale of the QED breakdown, $\Lambda$,
\cite{bib:drell,bib:low}. This is the scale up to which the $e\gamma$ interaction can be 
described as point-like. The deviation cross-section would be given 
by expression~(\ref{eq:cutoff}) (where $\alpha$ is the fine-structure constant) 
which allows for negative and positive 
interference in the form of the $\pm 1/\Lambda_\pm^4$ parameterisation.

\begin{equation}
\biggl (\frac {d\sigma}{d\Omega}\biggr)^D =\pm \frac{\alpha^2 s}{2}(1+\scos \theta^\ast)\frac{1}{{\Lambda_\pm}^4} 
\label{eq:cutoff}
\end{equation}

The result of the maximum likelihood fit to data yielded 
$-59^{+40}_{-39}$~TeV$^{-4}$ for the estimator 
$1/\Lambda_\pm^4$, resulting in lower bounds for the energy scale of the QED breakdown of 379~GeV and 300~GeV for $\Lambda_+$ and $\Lambda_-$, respectively.

\subsection{Search for contact interactions}
 
Bounds on the mass scale of the $e^+e^-\gamma \gamma$ contact interaction  can be 
parameterised in different ways, depending on the dimensionality of the effective Lagrangian 
describing the interaction~\cite{bib:eboli}. 
Operators of dimension 6, 7 and 8 translate into the characteristic scales $\Lambda_6$, $\Lambda_7$ and $\Lambda_8$. The corresponding deviation cross-sections can be expressed as:

\begin{align}
&\biggl (\frac {d\sigma}{d\Omega}\biggr)^D = 
\; \alpha s(1+\scos \theta^\ast)\frac{1}{{\Lambda_6}^4} \quad , \\
&\biggl (\frac {d\sigma}{d\Omega}\biggr)^D = 
\; \frac{s^2}{32\pi}\frac{1}{{\Lambda_7}^6}\quad , \\
&\biggl (\frac {d\sigma}{d\Omega}\biggr)^D = 
\; \frac{s^2 M_e^2}{64\pi}\frac{1}{{\Lambda_8}^8}\quad ,
\label{eq:dsdt-contact}
\end{align}

\noindent 
where $M_e$ is the electron mass.

Maximum likelihood fits to data yielded the results 
$1/\Lambda_6^4=-0.22^{+0.15}_{-0.14}$~TeV$^{-4}$, 
$1/\Lambda_7^6=-4.1^{+3.2}_{-3.1}$~TeV$^{-6}$ and 
$1/\Lambda_8^8=(-31\pm{24})\times 10^{12}$~TeV$^{-8}$, resulting 
in 95\%~C.L. lower bounds on the characteristic energy scales for contact interactions, 
$\Lambda_6$, $\Lambda_7$ and $\Lambda_8$ of 
1.5~TeV, 790~GeV and 21~GeV respectively.

\subsection{Search for excited electrons}

Within the framework of composite models,
deviations from QED could 
also follow from the t-channel exchange of an excited electron. 
In the case of an $e^\ast e \gamma$ 
chiral magnetic coupling~\cite{bib:vachon},
the deviation from the QED differential cross-section is expressed by:

\begin{equation}
\biggl (\frac {d\sigma}{d\Omega}\biggr)^D =
\frac{\alpha^2}{4} \biggl( \frac{f_\gamma}{\Lambda} \biggr)^2
\sum_{n=1}^2 
\left[
\frac{(1+i^{2n}\cos\theta^\ast)^2}{Y+i^{2n}\cos\theta^\ast} 
 \biggl(1+\frac{s}{4}\biggl( \frac{f_\gamma}{\Lambda} \biggr)^2
 \frac{1-\scos\theta^\ast}{Y+i^{2n}\cos\theta^\ast} \biggr)
\right]
\label{eq:mex}
\end{equation}


\noindent
where  $i$ is the imaginary number and Y=1+$2M_{e^\ast}^2/s$, $M_{e^\ast}$ being the mass of the excited electron.
$f_\gamma/\Lambda$ is the coupling constant with $f_\gamma=-\frac{1}{2}(f+f^{'})$, 
where $f$ and $f^{'}$ are weight factors associated to the different gauge groups.

\begin{figure}[H]
\begin{center}
\scalebox{0.6}{\includegraphics[bb=13 27 524 430]{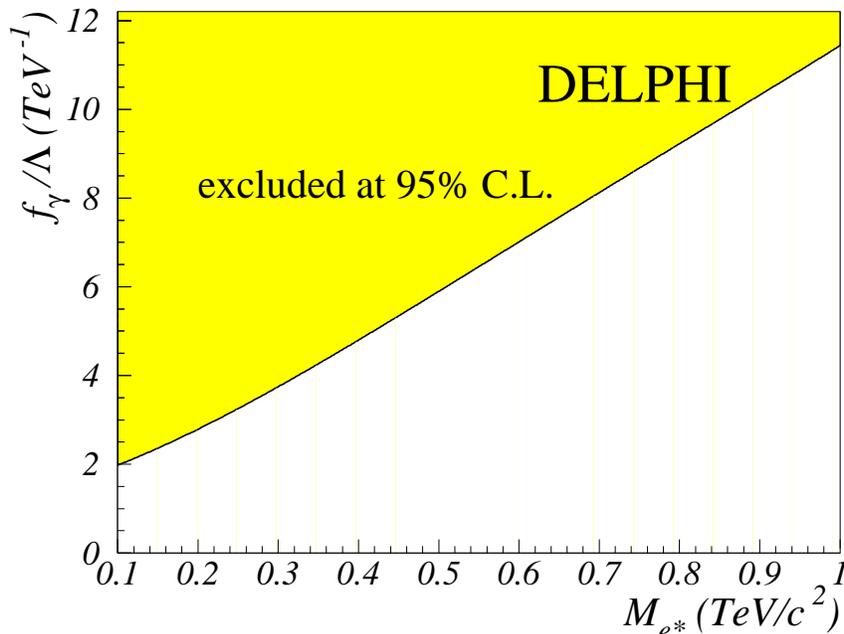}}
\end{center}
\vspace*{-0.3cm}
\caption{95\% C.L. upper bound on the coupling constant $f/\Lambda$ (for $f_\gamma=f=f^{'}$)
as a function of the mass of an excited electron with a chiral magnetic 
coupling to the photon-electron pair.
\label{fig:fl.vs.me}}
\end{figure}

In order to derive a lower bound on the excited electron mass,
$f_\gamma/\Lambda$ was set to $1/M_{e^\ast}$ in expression~(\ref{eq:mex}).
A 95\%~C.L. lower bound $M_{e^\ast}$ = 295~GeV/$c^2$ was derived, corresponding to 
$1/M_{e^\ast}^4=-142^{+104}_{-113}$~(TeV/c$^2$)$^{-4}$.
In addition, limits on the coupling constant $f_\gamma/\Lambda$ as a function of $M_{e^\ast}$ were derived by performing a scan over 
$M_{e^\ast}$ and are presented in figure~\ref{fig:fl.vs.me}. 

The framework adopted in the interpretation of the previous DELPHI results 
\cite{bib:96-97,bib:98-99}, corresponding to 
a purely magnetic $e^\ast e \gamma$ coupling~\cite{bib:composite}, 
is strongly limited by $g_{e}-2$ measurements~\cite{bib:Renard,bib:Brodsky-Drell} 
for the energy scales accessible at LEP.
In such a framework, the maximum likelihood fit yielded 
$1/M_{e^\ast}=-69^{+49}_{-53}$~(TeV/c$^2$)$^{-4}$ 
and the 95\%~C.L. lower bound on $M_{e^\ast}$ was 356~GeV/c$^2$ 
(for $\lambda_\gamma$=1, where $f_\gamma/\Lambda= \sqrt{2}\lambda_\gamma/M_{e^\ast}$).

\subsection{Search for TeV-scale quantum gravity}

The phenomenological implications of large 
extra-dimensions~\cite{bib:giudice,bib:agashe}
have lead to the suggestion of the 
possibility of observing the effect of  
virtual graviton exchange at LEP as a departure of the 
differential cross-section for $e^+e^-\to \gamma\gamma$ 
from the QED prediction.
The deviation cross-section, given by expression~(\ref{eq:grav}),
can be parameterised as a function of the string mass scale $M_s$, which in some string 
models could be as low as the electroweak scale, and of a phase factor, $\lambda$, 
conventionally taken to be $\pm 1$~\footnote{The ratio $\lambda/M_s^4$ which follows the notation of~\cite{bib:hewett}
is related to the quantum gravity scale $\Lambda_T$ in Ref.~\cite{bib:giudice} as: 
$\frac{|\lambda|}{M_s^4}=-\pi/2(1/\Lambda_T^4)$.}.

\begin{equation}
\biggl (\frac {d\sigma}{d\Omega}\biggr)^D =
-\frac{\alpha s}{2\pi} (1+\scos\theta^\ast)\frac{\lambda}{M_s^4}
\label{eq:grav}
\end{equation}

A maximum likelihood fit yielded $\pm 1/M_s^4=1.36^{+0.92}_{-0.90}$~(TeV/c$^2$)$^{-4}$,
resulting in lower limits on the string mass scale of 771~GeV/$c^2$ and 985~GeV/$c^2$ 
for $\lambda=1$ and $\lambda=-1$, respectively.


\section{Summary}

The reaction \eeintogg(\fot) was studied using the LEP 2 data
collected with the DELPHI detector at centre-of-mass 
energies ranging from 161~GeV to 208~GeV,
corresponding to  a total integrated luminosity of 656.4~pb$^{-1}$.
The differential and total cross-sections for the process 
$e^+e^- \to \gamma \gamma$ were measured. 
Good agreement between the data and the QED prediction 
was found. 
The absence of a deviation from QED was used to set 
95\% C.L. lower limits on the parameters of models predicting deviations from QED.
The QED cut-off parameters $\Lambda_{+}$ and $\Lambda_{-}$ were found to be 
greater than 379~GeV and 300~GeV, respectively. 
Lower limits on the characteristic energy scales of $e^+e^- \gamma \gamma$ 
contact in\-ter\-actions, $\Lambda_6=1.5$~TeV, $\Lambda_7=790$~GeV and $\Lambda_8=21$~GeV were obtained.
A lower limit for the mass of an excited electron with a chiral magnetic coupling to 
photon-electron pairs, ${\rm M_{e^{\ast}} >}$~295~GeV/c$^2$, was obtained.  
The  possible contribution of virtual gravitons to the process 
$e^+e^- \to \gamma \gamma$ was probed, resulting in the bounds 
$M_s >$~771~GeV/c$^2$ and $M_s >$~985~GeV/c$^2$ 
for $\rm \lambda=1$ and $\rm \lambda=-1$ respectively (where $\rm \lambda$ 
is a phase factor in some quantum gravity models).

\subsection*{Acknowledgements}
\vskip 3 mm
 We are greatly indebted to our technical 
collaborators, to the members of the CERN-SL Division for the excellent 
performance of the LEP collider, and to the funding agencies for their
support in building and operating the DELPHI detector.\\
We acknowledge in particular the support of \\
Austrian Federal Ministry of Education, Science and Culture,
GZ 616.364/2-III/2a/98, \\
FNRS--FWO, Flanders Institute to encourage scientific and technological 
research in the industry (IWT), Belgium,  \\
FINEP, CNPq, CAPES, FUJB and FAPERJ, Brazil, \\
Czech Ministry of Industry and Trade, GA CR 202/99/1362,\\
Commission of the European Communities (DG XII), \\
Direction des Sciences de la Mati$\grave{\mbox{\rm e}}$re, CEA, France, \\
Bundesministerium f$\ddot{\mbox{\rm u}}$r Bildung, Wissenschaft, Forschung 
und Technologie, Germany,\\
General Secretariat for Research and Technology, Greece, \\
National Science Foundation (NWO) and Foundation for Research on Matter (FOM),
The Netherlands, \\
Norwegian Research Council,  \\
State Committee for Scientific Research, Poland, SPUB-M/CERN/PO3/DZ296/2000,
SPUB-M/CERN/PO3/DZ297/2000, 2P03B 104 19 and 2P03B 69 23(2002-2004)\\
FCT - Funda\c{c}\~ao para a Ci\^encia e Tecnologia, Portugal, \\
Vedecka grantova agentura MS SR, Slovakia, Nr. 95/5195/134, \\
Ministry of Science and Technology of the Republic of Slovenia, \\
CICYT, Spain, AEN99-0950 and AEN99-0761,  \\
The Swedish Natural Science Research Council,      \\
Particle Physics and Astronomy Research Council, UK, \\
Department of Energy, USA, DE-FG02-01ER41155. \\
EEC RTN contract HPRN-CT-00292-2002.\\

\pagebreak

\begin{table}[H]
\begin{center}
\begin{tabular}{c  c cc  c c c c c cccc c cc }
\hline
\multicolumn{1}{c}{ Year} & 
\multicolumn{1}{c}{  } &
\multicolumn{2}{c}{ 1996} & 
\multicolumn{1}{c}{  } &
\multicolumn{1}{c}{ 1997} & 
\multicolumn{1}{c}{  } &
\multicolumn{1}{c}{ 1998} & 
\multicolumn{1}{c}{  } &
\multicolumn{4}{c}{ 1999} & 
\multicolumn{1}{c}{  } &
\multicolumn{2}{c}{ 2000} \\
\hline
\multicolumn{1}{c}{$\sqrt{s}$ (GeV)} & 
\multicolumn{1}{c}{ $\;$ } &
\multicolumn{1}{c}{ 161.3} &
\multicolumn{1}{c}{ 172.0} &
\multicolumn{1}{c}{ $\;$ } &
\multicolumn{1}{c}{ 182.7} &
\multicolumn{1}{c}{ $\;$ } &
\multicolumn{1}{c}{ 188.6} &
\multicolumn{1}{c}{ $\;$ } &
\multicolumn{1}{c}{ 191.6} &
\multicolumn{1}{c}{ 195.5} &
\multicolumn{1}{c}{ 199.5} &
\multicolumn{1}{c}{ 201.6} &
\multicolumn{1}{c}{ $\;$ } &
\multicolumn{1}{c}{ 205.7} &
\multicolumn{1}{c}{ 206.3} \\
\hline
\multicolumn{1}{c}{${\cal{L}}$ (pb$^{-1}$)} &
\multicolumn{1}{c}{  } &
\multicolumn{1}{c}{ 8.4} &
\multicolumn{1}{c}{ 8.8} &
\multicolumn{1}{c}{  } &
\multicolumn{1}{c}{ 49.0}  &
\multicolumn{1}{c}{  } &
\multicolumn{1}{c}{ 152.6} &
\multicolumn{1}{c}{  } &
\multicolumn{1}{c}{ 25.1}  &
\multicolumn{1}{c}{ 75.9}  &
\multicolumn{1}{c}{ 82.5}  &
\multicolumn{1}{c}{ 40.0}  &
\multicolumn{1}{c}{  } &
\multicolumn{1}{c}{ 160.3} &
\multicolumn{1}{c}{ 54.0} \\
\hline
\end{tabular}
\end{center}
\vspace*{-0.5cm}
\caption
[LEP 2 data sets]
{Data sets used in the analysis, corresponding average 
centre-of-mass energies and integrated luminosities.
The data taken during the year 2000 were split in
two data sets, 
before and after an irreversible failure in a sector of the 
TPC. 
\label{tab:data}}
\end{table}


\vspace*{-0.5cm}

\begin{table}[H]
\begin{center}
\begin{tabular}{c r r r r r r }
\hline
 & & & & & & \\
\multicolumn{1}{c}{$\sqrt{s}$ (GeV)} &
\multicolumn{1}{c}{$N_{QED}$ } &
\multicolumn{1}{c}{$N_{Compton}$ } &
\multicolumn{1}{c}{$N_{Bhabha}$ } &
\multicolumn{1}{c}{$N_{\nu \bar{\nu}\gamma \gamma}$ } &
\multicolumn{1}{c}{$N_{SM}^{total}$} &
\multicolumn{1}{c}{$N_{data}$ } \\
 & & & & & & \\
\hline
161.3  &  69.34$\pm$ 1.99 & 0.99$\pm$0.91 & 0.04$\pm$0.02 & 0.43$\pm$0.04 &  70.8$\pm$ 2.2 &  76\\
172.0  &  64.23$\pm$ 1.85 & 0.91$\pm$0.84 & 0.03$\pm$0.02 & 0.46$\pm$0.04 &  65.6$\pm$ 2.0 &  47\\
182.7  & 316.97$\pm$ 9.11 & 3.57$\pm$2.03 & 0.06$\pm$0.02 & 2.54$\pm$0.23 & 323.1$\pm$ 9.3 & 304\\
188.6  & 892.97$\pm$ 4.81 & 7.99$\pm$0.50 & 0.07$\pm$0.05 & 7.91$\pm$0.73 & 908.9$\pm$ 4.9 & 888\\
191.6  & 147.87$\pm$ 1.27 & 1.37$\pm$0.25 & 0.07$\pm$0.01 & 1.55$\pm$0.17 & 150.9$\pm$ 1.3 & 133\\   
195.5  & 429.80$\pm$ 3.68 & 3.99$\pm$0.74 & 0.19$\pm$0.03 & 4.00$\pm$0.35 & 438.0$\pm$ 3.9 & 437\\   
199.5  & 453.86$\pm$ 3.91 & 4.16$\pm$0.77 & 0.20$\pm$0.03 & 4.39$\pm$0.33 & 462.6$\pm$ 4.0 & 434\\   
201.6  & 215.45$\pm$ 1.86 & 1.97$\pm$0.37 & 0.10$\pm$0.02 & 1.73$\pm$0.20 & 219.3$\pm$ 1.9 & 207\\   
205.7  & 835.04$\pm$ 7.20 &10.65$\pm$1.54 & 0.38$\pm$0.03 & 8.13$\pm$0.76 & 854.2$\pm$ 7.4 & 804\\   
206.3  & 275.72$\pm$ 3.16 & 3.55$\pm$0.75 & 0.12$\pm$0.03 & 2.47$\pm$0.28 & 281.9$\pm$ 3.3 & 274\\   
\hline 
total  &3701.29$\pm$14.48& 39.15$\pm$3.19& 1.23$\pm$0.09& 33.61$\pm$1.24 & 3775.3$\pm$14.9 & 3604\\  
\hline
\end{tabular}
\end{center}
\vspace*{-0.5cm}
\caption
{Number of events remaining in all real and simulated data 
samples after the application of 
the multi-photonic event selection criteria 
and before corrections for trigger and selection efficiency.
\label{tab:multi_data-sim}}
\end{table}

\vspace*{-0.5cm}


\begin{table}[H]
\begin{center}
\begin{tabular}{c r r c c r r}
\hline                                                                                
 & & & & & & \\
\multicolumn{1}{c}{\raisebox{-4pt}{$\sqrt{s}$ (GeV)}} &
\multicolumn{2}{c}{$<\varepsilon_{sel}^{\gamma\gamma}>$} & 
\multicolumn{2}{c}{$<\varepsilon_{trig}^{\gamma\gamma}>$} & 
\multicolumn{2}{c}{$N^{\gamma\gamma}$} \\
\multicolumn{1}{c}{ } &
\multicolumn{1}{c}{Barrel} &
\multicolumn{1}{c}{Endcaps} & 
\multicolumn{1}{c}{Barrel} &
\multicolumn{1}{c}{Endcaps} & 
\multicolumn{1}{c}{QED} &
\multicolumn{1}{c}{data} \\
 & & & & & & \\
\hline
161.3       & 0.770$\pm$0.020 & 0.519$\pm$0.026 &  0.97$\pm$0.01 & 1.000$^{+0.000}_{-0.003}$    &  51.5$\pm$ 2.8 &  57 \\
172.0       & 0.766$\pm$0.020 & 0.518$\pm$0.026 &  0.78$\pm$0.03 & 1.000$^{+0.000}_{-0.003}$    &  41.0$\pm$ 2.5 &  33 \\
182.7       & 0.773$\pm$0.015 & 0.522$\pm$0.025 & 0.977$\pm$0.005& 0.998$\pm$0.001             & 234.1$\pm$ 8.0 & 220 \\
188.6       & 0.766$\pm$0.007 & 0.536$\pm$0.011 & 0.984$\pm$0.002& 0.9998$\pm$0.0002            & 666.5$\pm$ 6.8 & 673 \\
191.6       & 0.791$\pm$0.007 & 0.615$\pm$0.014 & 0.979$\pm$0.007& 1.000$^{+0.000}_{-0.001}$    & 114.6$\pm$ 1.9 & 102 \\
195.5       & 0.777$\pm$0.006 & 0.618$\pm$0.013 & 0.976$\pm$0.004& 1.0000$^{+0.0000}_{-0.0003}$ & 327.3$\pm$ 3.8 & 341 \\
199.5       & 0.772$\pm$0.007 & 0.609$\pm$0.013 & 0.963$\pm$0.005& 0.9993$\pm$0.0005            & 336.6$\pm$ 4.1 & 325 \\
201.6       & 0.783$\pm$0.008 & 0.622$\pm$0.013 & 0.983$\pm$0.005& 1.0000$^{+0.0000}_{-0.0007}$ & 162.5$\pm$ 2.6 & 150 \\
205.7       & 0.775$\pm$0.006 & 0.608$\pm$0.013 & 0.976$\pm$0.003& 0.9989$\pm$0.0004            & 620.7$\pm$ 7.0 & 575 \\ 
206.3       & 0.781$\pm$0.012 & 0.589$\pm$0.015 & 0.964$\pm$0.006& 0.9989$\pm$0.0008            & 206.4$\pm$ 3.9 & 203 \\
\hline
total & & & & & 2761.2$\pm$15.2 & 2679  \\
\hline
\end{tabular}
\end{center}
\vspace{-0.5cm}
\caption{Average selection efficiency (with full uncertainties), 
and average trigger efficiency for $\gamma \gamma$ (with statistical uncertainties) 
in the barrel and endcaps of DELPHI, number of events expected from simulation, 
including selection and trigger efficiency corrections and total number of 
$e^+e^-\to\gamma \gamma$ events selected from each of the ten data samples.
\label{tab:Nqed}}
\end{table}

\begin{table}[H]
\begin{center}
\begin{tabular}{l c r | l c r}
\hline
\multicolumn{1}{c}{\raisebox{1pt}{$|\cos{ \theta^\ast}|$}} &
\multicolumn{1}{c}{ } & 
\multicolumn{1}{c|}{R} & 
\multicolumn{1}{c}{\raisebox{1pt}{$|\cos{ \theta^\ast}|$}} &
\multicolumn{1}{c}{ } & 
\multicolumn{1}{c}{R } \\
\hline
0.0350-0.0855   &   &1.071 $\pm$ 0.008  &                              &    &\\ 
0.0855-0.1360   &   &1.066 $\pm$ 0.008  & {\raisebox{4pt}{0.05-0.10}}  &    &{\raisebox{4pt}{1.070 $\pm$ 0.008}} \\ 
0.1360-0.1865   &   &1.065 $\pm$ 0.008  & {\raisebox{4pt}{0.10-0.15}}  &    &{\raisebox{4pt}{1.066 $\pm$ 0.008}} \\
0.1865-0.2370   &   &1.070 $\pm$ 0.008  & {\raisebox{4pt}{0.15-0.20}}  &    &{\raisebox{4pt}{1.066 $\pm$ 0.008}} \\ 
0.2370-0.2875   &   &1.068 $\pm$ 0.007  & {\raisebox{4pt}{0.20-0.25}}  &    &{\raisebox{4pt}{1.070 $\pm$ 0.008}} \\ 
0.2875-0.3380   &   &1.056 $\pm$ 0.007  & {\raisebox{4pt}{0.25-0.30}}  &    &{\raisebox{4pt}{1.065 $\pm$ 0.008}} \\ 
0.3380-0.3885   &   &1.057 $\pm$ 0.007  & {\raisebox{4pt}{0.30-0.35}}  &    &{\raisebox{4pt}{1.057 $\pm$ 0.007}} \\ 
0.3885-0.4390   &   &1.048 $\pm$ 0.007  & {\raisebox{4pt}{0.35-0.40}}  &    &{\raisebox{4pt}{1.055 $\pm$ 0.007}}\\ 
0.4390-0.4895   &   &1.054 $\pm$ 0.006  & {\raisebox{4pt}{0.40-0.45}}  &    &{\raisebox{4pt}{1.050 $\pm$ 0.007}} \\ 
0.4895-0.5400   &   &1.059 $\pm$ 0.006  & {\raisebox{4pt}{0.45-0-50}}  &    &{\raisebox{4pt}{1.055 $\pm$ 0.006}} \\ 
0.5400-0.5905   &   &1.042 $\pm$ 0.006  & {\raisebox{4pt}{0.50-0.55}}  &    &{\raisebox{4pt}{1.056 $\pm$ 0.006}} \\ 
0.5905-0.6410   &   &1.050 $\pm$ 0.005  & {\raisebox{4pt}{0.55-0.60}}  &    &{\raisebox{4pt}{1.044 $\pm$ 0.007}} \\ 
0.6410-0.6860   &   &1.031 $\pm$ 0.005  & {\raisebox{4pt}{0.60-0.65}}  &    &{\raisebox{4pt}{1.047 $\pm$ 0.005}} \\ 
0.6860-0.7310   &   &1.049 $\pm$ 0.005  & {\raisebox{4pt}{0.65-0.70}}  &    &{\raisebox{4pt}{1.036 $\pm$ 0.005}} \\ 
{\raisebox{-2pt}{0.8190-0.8625}}&   & {\raisebox{-2pt}{1.037 $\pm$ 0.004}}  &   &      \\
{\raisebox{-4pt}{0.8625-0.9060}}&   & {\raisebox{-4pt}{1.035 $\pm$ 0.003}}  & {\raisebox{2pt}{0.85-0.90}} &&{\raisebox{2pt}{1.035 $\pm$ 0.003}} \\
\hline	         
\end{tabular}
\end{center}
\vspace*{-0.5cm}
\caption{
${\cal O}(\alpha^3)$ 
radiative correction factors, R, evaluated from the generator 
of Berends and Kleiss~[2] as a function of $|\cos \theta^\ast|$ for 
the binning used in the present analysis and for the binning chosen in order to combine the data of
the four LEP collaborations. The dependence of the radiative correction factor with $\sqrt{s}$ 
is negligible.\label{tab:radcor}}
\end{table}

\begin{table}[H]
\begin{center}
\begin{tabular}{c c c c c c c c c c cc cc  c}
\hline
\multicolumn{1}{c}{$\sqrt{s}$} &
\multicolumn{1}{c}{ } &
\multicolumn{13}{c}{Systematic uncertainties (\%)} \\
\cline{2-15}
\multicolumn{1}{c}{(GeV)} &
\multicolumn{1}{c}{ } &
\multicolumn{1}{c}{Bkg} &
\multicolumn{1}{c}{ } &
\multicolumn{1}{c}{$\epsilon_{sel}$} &
\multicolumn{1}{c}{ } &
\multicolumn{1}{c}{${\cal{R}}$} &
\multicolumn{1}{c}{ } &
\multicolumn{1}{c}{$\epsilon_{trig}$} &
\multicolumn{1}{c}{ } &
\multicolumn{1}{c}{$R$} &
\multicolumn{1}{c}{ } &
\multicolumn{1}{c}{${\cal{L}}$} &
\multicolumn{1}{c}{ } &
\multicolumn{1}{c}{Total} \\
\hline
 161.3 & {\hspace{.2cm}}   &0.10  &{\hspace{.2cm}}  & 1.94 & 
{\hspace{.2cm}}   &1.20& {\hspace{.2cm}}  & 0.60 & {\hspace{.2cm}}  
&0.13 & {\hspace{.2cm}}   &0.56 &{\hspace{.2cm}}    &2.43\\
 172.0 &  &0.10  &  & 1.94 &  &1.24& & 1.80 &  &0.13 &  &0.56 &  &2.98\\
 182.7 &  &0.06  &  & 1.94 &  &0.49& & 0.23 &  &0.13 &  &1.03 &  &2.27\\
 188.6 &  &0.08  &  & 0.72 &  &0.48& & 0.11 &  &0.13 &  &0.56 &  &1.05\\
 191.6 &  &0.15  &  & 0.84 &  &0.55& & 0.30 &  &0.13 &  &0.56 &  &1.20\\
 195.5 &  &0.14  &  & 0.84 &  &0.35& & 0.18 &  &0.13 &  &0.56 &  &1.10\\
 199.5 &  &0.11  &  & 0.82 &  &0.41& & 0.24 &  &0.13 &  &0.56 &  &1.11\\
 201.6 &  &0.11  &  & 0.82 &  &0.51& & 0.22 &  &0.13 &  &0.56 &  &1.15\\
 205.7 &  &0.09  &  & 0.83 &  &0.32& & 0.14 &  &0.13 &  &0.56 &  &1.07\\
 206.3 &  &0.20  &  & 0.95 &  &0.71& & 0.31 &  &0.13 &  &0.56 &  &1.37 \\  
\hline
\end{tabular}
\end{center}
\vspace*{-0.5cm}
\caption
{Systematic uncertainties on the cross-section measurements computed for the analysed data sets.
A data taking instability led to the assignment of a $\pm$1\% experimental error 
to the 182.7~GeV data luminosity (the theoretical error is of $\pm$0.25\%).
\label{tab:esyst}}
\end{table}


\begin{table}[H]
\begin{center}
\begin{tabular}{c r r r r r}  
\hline
\multicolumn{6}{c}{ }  \\
\multicolumn{1}{c}{\raisebox{1pt}{$|\cos{ \theta^\ast}|$}} &
\multicolumn{5}{c}{\raisebox{3pt}{ $N^{\gamma \gamma}_{data}/\varepsilon^{\gamma\gamma}$}} \\
              & 161.3~GeV &172.0~GeV &182.7~GeV  & 188.6~GeV &191.6~GeV \\
\cline{2-6}
0.0350-0.0855 &  3/ 0.751&  1/ 0.751 & 5/ 0.653  & 18/ 0.759 & 3/ 0.832 \\         
0.0855-0.1360 &  0/ 0.816&  1/ 0.816 & 6/ 0.641  & 33/ 0.707 & 3/ 0.802 \\         
0.1360-0.1865 &  2/ 0.973&  2/ 0.730 &13/ 0.895  & 28/ 0.761 & 3/ 0.776 \\         
0.1865-0.2370 &  2/ 0.840&  2/ 0.420 &11/ 0.708  & 28/ 0.744 & 2/ 0.776 \\         
0.2370-0.2875 &  3/ 0.912&  0/ 0.608 &10/ 0.839  & 30/ 0.734 & 4/ 0.847 \\         
0.2875-0.3380 &  2/ 0.833&  3/ 0.570 & 6/ 0.830  & 33/ 0.696 & 4/ 0.820 \\         
0.3380-0.3885 &  3/ 0.584&  0/ 0.497 & 5/ 0.761  & 26/ 0.658 & 4/ 0.837 \\         
0.3885-0.4390 &  0/ 0.524&  0/ 0.458 & 4/ 0.617  & 29/ 0.705 & 2/ 0.800 \\         	   
0.4390-0.4895 &  5/ 0.735&  0/ 0.680 &11/ 0.752  & 32/ 0.772 & 6/ 0.748 \\ 
0.4895-0.5400 &  3/ 0.895&  4/ 0.716 &11/ 0.836  & 39/ 0.814 & 5/ 0.784 \\ 
0.5400-0.5905 &  2/ 0.647&  3/ 0.543 &14/ 0.733  & 42/ 0.802 & 6/ 0.808 \\ 
0.5905-0.6410 &  5/ 0.669&  2/ 0.606 &19/ 0.719  & 38/ 0.776 &10/ 0.798 \\ 
0.6410-0.6860 &  2/ 0.832&  2/ 0.667 &21/ 0.769  & 55/ 0.739 & 9/ 0.734 \\ 
0.6860-0.7310 &  5/ 0.740&  3/ 0.514 &23/ 0.750  & 57/ 0.738 & 1/ 0.698 \\ 
0.8190-0.8625 &  5/ 0.581&  6/ 0.592 &24/ 0.579  & 80/ 0.596 &18/ 0.629 \\ 
0.8625-0.9060 & 15/ 0.472&  4/ 0.456 &37/ 0.478  &105/ 0.495 &22/ 0.603 \\         
\hline	                                	  	                                  
\end{tabular}	
\vspace*{0.2cm}
\begin{tabular}{c r r r r r}  
\hline
\multicolumn{6}{c}{ }  \\
\multicolumn{1}{c}{\raisebox{1pt}{$|\cos{ \theta^\ast}|$}} &
\multicolumn{5}{c}{\raisebox{3pt}{ $N^{\gamma \gamma}_{data}/\varepsilon^{\gamma\gamma}$}} \\
          & 195.5~GeV &199.5~GeV  & 201.6~GeV  & 205.7~GeV & 206.3~GeV \\
\cline{2-6}
0.0350-0.0855 &   9/ 0.762 &  11/ 0.745 &   6/ 0.754 &   20/0.695 &   6/ 0.760\\         
0.0855-0.1360 &  10/ 0.742 &   8/ 0.711 &   7/ 0.726 &   17/0.803 &   6/ 0.777\\         
0.1360-0.1865 &  21/ 0.740 &   9/ 0.802 &   4/ 0.736 &   23/0.703 &   6/ 0.889\\         
0.1865-0.2370 &  10/ 0.821 &  12/ 0.831 &   2/ 0.853 &   30/0.806 &   9/ 0.838\\         
0.2370-0.2875 &   7/ 0.768 &  16/ 0.744 &   9/ 0.823 &   17/0.778 &   8/ 0.930\\         
0.2875-0.3380 &   6/ 0.738 &  11/ 0.744 &   7/ 0.730 &   19/0.781 &  10/ 0.854\\         
0.3380-0.3885 &  12/ 0.762 &  18/ 0.736 &   5/ 0.813 &   29/0.698 &  10/ 0.523\\         
0.3885-0.4390 &  15/ 0.726 &  16/ 0.638 &   7/ 0.608 &   24/0.762 &  11/ 0.619\\         
0.4390-0.4895 &  15/ 0.821 &  19/ 0.803 &   9/ 0.804 &   17/0.797 &   8/ 0.886\\         
0.4895-0.5400 &  29/ 0.782 &  21/ 0.745 &   6/ 0.754 &   32/0.763 &   9/ 0.762\\         
0.5400-0.5905 &  19/ 0.810 &  25/ 0.779 &  12/ 0.797 &   36/0.798 &   9/ 0.818\\         
0.5905-0.6410 &  28/ 0.789 &  25/ 0.729 &  11/ 0.774 &   34/0.718 &  16/ 0.762\\         
0.6410-0.6860 &  25/ 0.709 &  10/ 0.736 &   9/ 0.736 &   52/0.746 &  12/ 0.732\\         
0.6860-0.7310 &  28/ 0.716 &  24/ 0.743 &  14/ 0.764 &   51/0.754 &  21/ 0.712\\         
0.8190-0.8625   &  44/ 0.642 &  37/ 0.637 &  19/ 0.653 &   76/0.644 &  33/ 0.630\\         
0.8625-0.9060   &  63/ 0.601 &  63/ 0.598 &  23/ 0.600 &   98/0.583 &  29/ 0.560\\         
\hline	                               	  	                              
\end{tabular}
\end{center}
\vspace{-0.5cm}
\caption
{Number of selected data events and efficiencies (selection times trigger
    efficiencies corrected for the residual discrepancies between real data
    and simulation)
as a function of $|\cos{\theta^\ast}|$ and $\sqrt{s}$ 
within the polar angle range covered by DELPHI's electromagnetic 
calorimeters:
$|\cos{\theta^\ast}|\in [0.035,0.731] \cup [0.819,0.906]$.
\label{tab:nev}}
\end{table}

\begin{table}[H]
\begin{center}
\begin{tabular}{c r r r r r}  
\hline
\multicolumn{6}{c}{ }  \\
\multicolumn{1}{c}{\raisebox{1pt}{$|\cos{ \theta^\ast}|$}} &
\multicolumn{5}{c}{\raisebox{3pt}{ $N^{\gamma \gamma}_{data}/\varepsilon^{\gamma\gamma}$}} \\
          & 161.3~GeV &172.0~GeV &182.7~GeV  & 188.6~GeV &191.6~GeV \\
\cline{2-6}
0.05-0.10 &  2/0.876  &1/0.876   &  4/0.775  & 19/0.773  &  4/0.834  \\         
0.10-0.15 &  0/0.670  &1/0.595   &  7/0.526  & 38/0.714  &  2/0.803  \\         
0.15-0.20 &  2/1.000  &3/0.711   & 13/0.964  & 23/0.764  &  4/0.773  \\         
0.20-0.25 &  3/0.790  &1/0.429   & 10/0.672  & 29/0.731  &  3/0.799  \\         
0.25-0.30 &  2/0.878  &3/0.624   &  9/0.828  & 28/0.730  &  5/0.819  \\         
0.30-0.35 &  2/0.792  &0/0.549   &  8/0.807  & 32/0.707  &  2/0.844  \\         
0.35-0.40 &  3/0.595  &0/0.481   &  2/0.732  & 24/0.661  &  4/0.840  \\         	   
0.40-0.45 &  0/0.553  &0/0.496   &  5/0.656  & 29/0.705  &  3/0.772  \\ 
0.45-0-50 &  6/0.743  &1/0.705   & 12/0.758  & 41/0.773  &  5/0.763  \\ 
0.50-0.55 &  2/0.790  &4/0.603   & 11/0.744  & 39/0.812  &  4/0.792  \\ 
0.55-0.60 &  3/0.705  &2/0.627   & 15/0.802  & 41/0.802  &  8/0.797  \\ 
0.60-0.65 &  4/0.668  &2/0.593   & 17/0.723  & 31/0.759  & 11/0.785  \\ 
0.65-0.70 &  3/0.823  &3/0.630   & 25/0.746  & 65/0.749  &  6/0.745  \\ 
0.85-0.90 & 16/0.516  &5/0.526   & 41/0.514  &110/0.521  & 22/0.632  \\ 
\hline	                               	  	                                   	     \end{tabular}
\vspace*{0.2cm}
\begin{tabular}{c r r r r r}  
\hline
\multicolumn{6}{c}{ }  \\
\multicolumn{1}{c}{\raisebox{1pt}{$|\cos{ \theta^\ast}|$}} &
\multicolumn{5}{c}{\raisebox{3pt}{ $N^{\gamma \gamma}_{data}/\varepsilon^{\gamma\gamma}$}} \\
          & 195.5~GeV &199.5~GeV  & 201.6~GeV  & 205.7~GeV & 206.3~GeV \\
\cline{2-6}
0.05-0.10 & 8/0.769   &11/0.738   &  7/0.766  & 23/0.703  &  3/0.804  \\         
0.10-0.15 &12/0.770   & 6/0.733   &  6/0.723  & 20/0.816  &  9/0.729  \\         
0.15-0.20 &24/0.717   &11/0.794   &  4/0.733  & 16/0.700  &  4/0.913  \\         
0.20-0.25 & 5/0.844   &13/0.848   &  6/0.875  & 31/0.794  & 12/0.888  \\         
0.25-0.30 & 8/0.749   &15/0.746   &  5/0.826  & 17/0.803  &  6/0.902  \\         
0.30-0.35 & 7/0.751   &13/0.747   &  7/0.734  & 20/0.750  & 14/0.833  \\         
0.35-0.40 &15/0.763   &15/0.677   &  5/0.769  & 28/0.709  &  4/0.493  \\     
0.40-0.45 &11/0.716   &18/0.676   &  7/0.625  & 24/0.776  & 13/0.654  \\ 
0.45-0-50 &18/0.834   &20/0.802   &  9/0.802  & 17/0.784  &  7/0.877  \\ 
0.50-0.55 &29/0.792   &20/0.738   &  7/0.749  & 32/0.766  &  9/0.806  \\ 
0.55-0.60 &20/0.792   &25/0.781   & 11/0.799  & 35/0.804  & 14/0.788  \\ 
0.60-0.65 &25/0.775   &21/0.718   & 11/0.759  & 33/0.728  & 11/0.746  \\ 
0.65-0.70 &27/0.723   &16/0.758   & 10/0.764  & 66/0.755  & 16/0.746  \\ 
0.85-0.90 &76/0.645   &67/0.651   & 32/0.667  &122/0.657  & 32/0.628  \\ 
\hline	                               	  	                                   	         
\end{tabular}
\end{center}
\vspace{-0.5cm}
\caption
{Number of data events within the 
range $|\cos{\theta^\ast}|\in [0.05,0.70] \cup [0.85,0.90]$,   
as a function of $|\cos{\theta^\ast}|$ and $\sqrt{s}$ and corresponding 
efficiencies (selection times trigger
    efficiencies corrected for the residual discrepancies between real data
    and simulation). 
The binning used was chosen in order to combine the data analysed by the four LEP collaborations.
\label{tab:lwg}}
\end{table}


\begin{table}[H]
\begin{center}
\begin{tabular}{c c r c }
\hline
\multicolumn{2}{c}{ } & 
\multicolumn{1}{c}{ $ d\sigma^{0}/d\Omega$}   &
\multicolumn{1}{c}{ $ d\sigma^{0}_{QED}/d\Omega$}   \\ 
\multicolumn{1}{c}{\raisebox{4.pt}{ $|\cos\theta^\ast|$}}    & 
\multicolumn{1}{c}{ } & 
\multicolumn{1}{c}{\raisebox{2.pt}{ (pb/str)}}   &
\multicolumn{1}{c}{\raisebox{2.pt}{ (pb/str)}}   \\ 
\hline
0.0350-0.0855&\hspace{0.5cm}  &   0.50 $\pm$ 0.05 $\pm$ 0.02  &   0.55  \\ 
0.0855-0.1360&  &   0.56 $\pm$ 0.06 $\pm$ 0.02  &   0.56  \\ 
0.1360-0.1865&  &   0.69 $\pm$ 0.06 $\pm$ 0.02  &   0.57  \\ 
0.1865-0.2370&  &   0.63 $\pm$ 0.06 $\pm$ 0.03  &   0.59  \\ 
0.2370-0.2875&  &   0.60 $\pm$ 0.06 $\pm$ 0.02  &   0.62  \\ 
0.2875-0.3380&  &   0.62 $\pm$ 0.06 $\pm$ 0.02  &   0.66  \\ 
0.3380-0.3885&  &   0.74 $\pm$ 0.07 $\pm$ 0.02  &   0.71  \\ 
0.3885-0.4390&  &   0.72 $\pm$ 0.07 $\pm$ 0.02  &   0.77  \\ 
0.4390-0.4895&  &   0.70 $\pm$ 0.06 $\pm$ 0.02  &   0.84  \\ 
0.4895-0.5400&  &   0.92 $\pm$ 0.07 $\pm$ 0.02  &   0.93  \\ 
0.5400-0.5905&  &   1.01 $\pm$ 0.08 $\pm$ 0.02  &   1.05  \\ 
0.5905-0.6410&  &   1.15 $\pm$ 0.08 $\pm$ 0.02  &   1.21  \\ 
0.6410-0.6860&  &   1.39 $\pm$ 0.10 $\pm$ 0.03  &   1.40  \\ 
0.6860-0.7310&  &   1.59 $\pm$ 0.11 $\pm$ 0.03  &   1.64  \\ 
0.8190-0.8625&  &   2.95 $\pm$ 0.16 $\pm$ 0.05  &   3.18  \\ 
0.8625-0.9060&  &   4.50 $\pm$ 0.21 $\pm$ 0.08  &   4.48  \\ 
\hline
\end{tabular}
\end{center}
\vspace*{-0.5cm}
\caption
{The differential cross-section for \eeintogg\ with statistical and 
systematic uncertainties,   
obtained by combining the ten data sets,  compared to the theoretical 
predictions from QED for each $|\cos{\theta^\ast}|$ interval.
The displayed values correspond to an average centre-of-mass 
energy of 195.6 GeV and to a total integrated luminosity of 656.4~pb$^{-1}$.
\label{tab:dsdo}}
\end{table}

\begin{table}[H]
\begin{center}
\begin{tabular}{c c r c r c c}
\hline
\multicolumn{1}{c}{\raisebox{-2.pt}{$\sqrt{s}$}} &
\multicolumn{1}{c}{ } &
\multicolumn{1}{c}{\raisebox{-2.pt}{$\sigma^0$}} &
\multicolumn{1}{c}{ } &
\multicolumn{1}{c}{\raisebox{-2.pt}{$\sigma^0_{QED}$}} & 
\multicolumn{2}{c}{ } \\
\multicolumn{1}{c}{\raisebox{3.pt}{(GeV)}} &
\multicolumn{1}{c}{ } &
\multicolumn{1}{c}{\raisebox{3.pt}{(pb)}} &
\multicolumn{1}{c}{ } &
\multicolumn{1}{c}{\raisebox{3.pt}{(pb)}} & 
\multicolumn{1}{c}{ } &
\multicolumn{1}{c}{\raisebox{3.pt}{$\chi^2/ndof$}} \\
\hline
161.3 &\hspace{0.2cm} &10.71 $\pm$ 1.42 $\pm$ 0.26 & &8.57 &&1.0\\
172.0 &               & 6.53 $\pm$ 1.14 $\pm$ 0.19 & &7.53 &&0.7\\
182.7 &               & 6.72 $\pm$ 0.45 $\pm$ 0.15 & &6.68 &&1.2\\ 
188.6 &               & 6.57 $\pm$ 0.25 $\pm$ 0.07 & &6.26 &&0.9\\
191.6 &               & 5.73 $\pm$ 0.57 $\pm$ 0.07 & &6.07 &&2.3\\
195.5 &               & 6.34 $\pm$ 0.34 $\pm$ 0.07 & &5.83 &&1.6\\
199.5 &               & 5.64 $\pm$ 0.31 $\pm$ 0.06 & &5.60 &&2.3\\
201.6 &               & 5.25 $\pm$ 0.43 $\pm$ 0.06 & &5.48 &&0.9\\
205.7 &               & 5.11 $\pm$ 0.21 $\pm$ 0.05 & &5.27 &&1.3\\
206.3 &               & 5.43 $\pm$ 0.38 $\pm$ 0.07 & &5.24 &&1.0\\
\hline                                    
195.6 &               & 5.66 $\pm$ 0.11 $\pm$ 0.03 & &5.83 &&1.3\\
\hline
\end{tabular}
\end{center}
\vspace*{-0.5cm}
\caption
{The visible Born level cross-section (with statistical and systematic uncertainties)
in the polar angle interval 
$\theta^{\ast} \in [25^\circ,35^\circ] \cup [43^\circ,88^\circ]$ and its complement
with respect to 90 degrees (corresponding to $|\cos{\theta^\ast}| \in [0.035,0.731] \cup [0.819,0.906]$),
for the ten data sets compared to the respective QED predictions 
and $\chi^{2} / ndof $ of the  
differential cross-section distributions with respect to the QED prediction.
The visible cross-section  corresponding to the combination of all 
data sets at an average centre-of-mass energy of 195.6~GeV and its respective theoretical 
prediction are also presented. 
\label{tab:stotqed}}
\end{table}


\begin{table}[hbt]
{\normalsize
\begin{center}
\begin{tabular}{c c r c l}
\hline
\multicolumn{1}{c}{  } &        
\multicolumn{1}{c}{\raisebox{-4pt}{ $\xi$ }} &        
\multicolumn{1}{c}{\raisebox{-4pt}{ $\xi^{+ \sigma_+}_{-\sigma_-}$}} &        
\multicolumn{1}{c}{\raisebox{-3pt}{Fit}}  &
\multicolumn{1}{c}{\raisebox{-3pt}{95\% C.L.}} \\        
\multicolumn{1}{c}{ } &        	
\multicolumn{1}{c}{ } &        
\multicolumn{1}{c}{ } &        
\multicolumn{1}{c}{parameter } &        
\multicolumn{1}{c}{limits} \\        
\hline
 & & & & \\
QED cutoff           & $\pm 1/\Lambda_\pm^4$ & 
\hspace{0.2cm} 
$-59.4^{+40.0}_{-39.0}$~{ TeV$^{-4 }$}  & 
$\Lambda_+$ & $\;$ 
379~{ GeV}\\
                     & & & 
$\Lambda_-$ & $\;$ 300~{ GeV}\\
 & & & & \\
Contact interactions &$1/\Lambda_6^4$ &
\hspace{0.2cm} 
$-0.22^{+0.15}_{-0.14}$~{ TeV$^{-4}$}   & 
$\Lambda_6$ & 
1537~{ GeV} \\
 & & & & \\
                     &$1/\Lambda_7^6$ & 
\hspace{0.2cm} 
$-4.1^{+3.2}_{-3.1} $~{ TeV$^{-6}$}   & 
$\Lambda_7$ & $\;$ 
790~{ GeV} \\
 & & & & \\
                     &$1/\Lambda_8^8$ & 
\hspace{0.2cm} 
$(-31\pm 24)\;10^{12}$~{ TeV$^{-8}$}  &
$\Lambda_8$ & $\;\;\;$ 21~{ GeV}  \\
 & & & & \\
$e^\ast$ exchange    & $1/M_{e^\ast}^4$ 
&\hspace{0.2cm} 
$-142^{+104}_{-113}$~{ (TeV/c$^2$)$^{-4}$}    &
$M_{e^\ast}$ & $\;$ 295~{ GeV/c$^2$}  \\ 
& & & & \\
Graviton exchange    & $\lambda/M_s^4$ & 
\hspace{0.2cm} $\; 1.36^{+0.92}_{-0.90}$~{ (TeV/c$^2$)$^{-4}$}  &  
${M_s}_{(\lambda=+1)}$ & $\;$ 771~{ GeV/c$^2$} \\
                     & & & 
${M_s}_{(\lambda=-1)}$ & $\;$ 
985~{ GeV/c$^2$} \\

                     & & & & \\

\hline
\end{tabular}
\end{center}
}
\vspace*{-0.5cm}
\caption{
Models predicting departures from QED and chosen estimators ($\rm \xi$).
The outputs of the likelihood function maximization
are presented in the third column whereas the  95\% C.L. lower limits 
on the free parameters of the models are presented in the fifth column.
The value obtained for $\rm \xi^{+\sigma_+}_{-\sigma_-}$ 
in case of 
the string mass scale corresponds to setting 
$|\lambda|$ to 1.\label{tab:devQED}}
\end{table}


\pagebreak

\pagebreak

\pagebreak


\vspace*{1.5cm}

\end{document}